\documentclass[preprint]{aastex}

\slugcomment{}

\shorttitle{Dust in Galaxies}
\shortauthors{Calzetti}

\begin{document}


\title{The Dust Opacity of Star-Forming Galaxies}

\author{Daniela Calzetti}
\affil{Space Telescope Science Institute, Baltimore, MD 21218}
\email{calzetti@stsci.edu}

\begin{abstract}
Presence of dust in galaxies removes half or more of the stellar energy
from the UV--optical budget of the Universe and has profound impact on our
understanding of how galaxies evolve. Measures of opacity in local galaxies
are reviewed together with widely used theoretical and empirical methods for
quantifying its effects. Existing evidence shows that the dust content of
nearby galaxies depends not only on their morphology, but also on their
luminosity and activity level. A digression is devoted to starbursts in view
of their potential relevance for measures of opacity in distant
galaxies. Scarcity of coherent multiwavelength datasets hampers our ability to
derive reliable obscuration estimates in intermediate and high redshift
galaxies. This, in turn, limits the reliability of inferred physical
quantities, such as star formation rates, stellar population ages, galaxy
luminosity functions, and others.
\end{abstract}


\tableofcontents

\section{Introduction}
After IRAS \citep{soi87} showed that most galaxies in the nearby Universe
contain at least some dust, the main question moved from ``whether'' to
``when, where, and how much''. ``When'' did galaxies started to be enriched in
metals and dust and how did their dust content evolve as a function of
redshift?  ``Where'' is the dust located within a galaxy and how is it
distributed among Hubble types? ``How much'' dust there is in galaxies, how
much stellar light is absorbed by dust and how is it re-emitted?

More than one third of the bolometric luminosity of local galaxies is
processed by dust into the infrared. The fraction of stellar radiation
reprocessed into the IRAS 8--120~$\mu$m window is $\sim$25\%--30\%
\citep{soi91}; this implies an infrared energy fraction around 35\%--40\% of
the total bolometric energy within the local $\sim$100~Mpc, once bolometric
corrections from the IRAS window to the full infrared range are included
\citep{des90,xu95,cal00}. This average value comes with a large variation
between galaxies, depending on luminosity, morphological type, and activity
level (see section~3).

The cosmic infrared background (CIB) observed by COBE
\citep{pug96,fix98,hau98} sets equally stringent limits on the total amount of
stellar energy re-processed by dust into the infrared at {\em all}
redshifts. About half, and possibly more, of the background radiation between
0.1~$\mu$m and 1000~$\mu$m is emerging at wavelengths $\lambda>$40~$\mu$m
\citep{hau01}. This implies that at some epoch(s) in the past galaxies have
been infrared--luminous at least as much as, and possibly more than, today to
produce the observed balance between UV/optical and infrared energy in the
cosmic background. Infrared--bright sources are also found to be about two
orders of magnitude more numerous in the past than today \citep[section~5.2
and][]{bar00}. The infrared luminosity of a galaxy is related to both its dust
content and nuclear/star-formation activity. The lower metal content of
distant galaxies does not necessarily make them more transparent systems than
galaxies today. The dust content of galaxies is, in fact, proportional to both
metallicity and gas content; as stellar populations evolve in time and more
gas is locked into stars, the metallicity of a galaxy increases while its gas
content decreases. The two trends are opposite to each other, and the dust
content reaches a peak sometimes during the evolution of the galaxy. Models
predict that galaxies at z$\sim$1--2 have about twice the dust opacity of
local galaxies \citep{wang91,edm97,cal99,pei99}. High-redshift, star-forming
galaxies are indirectly observed to have large gas column densities; the
average {\it SFR per unit galactic area} increases by more than an order of
magnitude between redshift 0 and 3 \citep{giaval96a,meu97,lanz99}, which is
indication of a similar increase in the average gas column density via the
global Schmidt law \citep{ken98}.

The presence of dust in galaxies limits our ability to interpret the local and
distant Universe, because dust extinction dims and redden the galaxy light in
the traditional UV--to--near-IR windows, where the vast majority of the data
have been obtained. Dust effects hamper the interpretation of galaxies SEDs in
terms of their fundamental parameters, such as age, stellar population mix,
star formation rates (SFRs), and stellar Initial Mass Function (IMF). Since
young stellar populations are on average more deeply embedded in dust clouds
than older stars, dust opacity variations within a galaxy impact the
derivation of a galaxy's star formation history, by altering the census of the
populations in the different age bins. Heavy dust extinction in nearby
starburst galaxies still prevents establishing whether the stellar IMF in
these systems is different from that of more quiescent environments
\citep{riek80,riek93,alon01}. The increase of dust content with luminosity in
spirals may be responsible for the non-linear behavior of the Tully-Fisher
relation, and this translates into biased distance estimates
\citep{gio95}. Determinations of luminosity functions, mass-to-light ratios,
and number counts, especially at short wavelengths, suffer from both the
presence of dust and the variations of dust content from galaxy to
galaxy. Light dimming by dust biases measurements of galaxies' SFRs, both
locally \citep{bel01,sul00} and at high redshift \citep{gla99,mad96,ste99},
and is the cause of uncertain cross-calibrations between different
indicators. The evolution with redshift of the SFR of galaxies per
unit comoving volume is still debated: unclear is whether it has a peak around
z$\sim$1--3 \citep{mad96,blain99} or is mostly flat between z$\sim$1 and
z$\sim$4 \citep{ste99,bar00}. Answering this question, which requires
understanding the dust opacity characteristics and the nature of high redshift
galaxies, can discriminate between different flavors of galaxy formation
scenarios \citep{whit91,orto95,baug98}. Finally, measuring the fraction of
metals locked into dust at different epochs is a crucial ingredient for
tracing the chemical evolution of galaxies.

The quantification of the dust opacity of galaxies of all types and at
all redshifts as a function of wavelength has required the collective
effort of many investigators over the last $\sim$10--15~years and is
still, in many aspects, an on-going process. This review attempts to 
capture the main results obtained so far; it concentrates on
star-forming galaxies, because of their key role for intepreting the
star formation history and evolution of galaxies. The topic has
acquired with time large breadth and numerous publications have
appeared in the field; despite the effort to include as many relevant
results as possible, some important contributions may have been
inadvertenly overlooked. 

After IRAS, the Infrared Space Observatory (ISO) continued to shed light on
the dust emission properties of galaxies, and comprehensive reviews are given
in \citet{gen00} and in the Proceedings of the 1998 Paris conference `The
Universe as Seen by ISO' \citep{cox99}. The status of the topic before ISO is
reported in the Proceedings of the 1994 Cardiff (U.K.) conference `The
Opacity of Spiral Disks' \citep{davie95} and of the 1996 Johannesburg (South
Africa) conference `New Extragalactic Perspectives in the New South Africa'
\citep{block96}.

Throughout the following, the adopted cosmology is a flat Universe with
$\Omega_{\Lambda}$=0 and Hubble constant
H$_o$=65~km~s$^{-1}$~Mpc$^{-1}$. `Local Universe' refers to a sphere of
100--200~Mpc radius centered on the Milky Way.

\section{The Radiative Transfer of Dust in Galaxies}

The transfer of radiation through dust is governed by the well-known 
equation \citep[e.g.,][]{cha60,sob63}:
\begin{equation}
{d I_{\lambda} \over d s} = -\kappa_{\lambda}
I_{\lambda}(s,\theta,\phi) + \epsilon_{\lambda}(s,\theta,\phi) +
\kappa_{\lambda} {\omega_{\lambda}\over
4\pi}\int_{\Omega}I_{\lambda}(s,\theta^{\prime},\phi^{\prime})
\Phi_{\lambda}(cos\,\Theta) d\Omega^{\prime},
\end{equation}
where I$_{\lambda}$(s,\,$\theta,\phi$) is the radiation intensity, ds is
the element of linear distance along the direction of propagation,
$d\Omega^{\prime}$ is the element of solid angle as seen by the dust
grain, $\kappa_{\lambda}$ is the opacity (absorption plus scattering)
of the dust per unit separation s, $\epsilon_{\lambda}(s,\theta,\phi)$
is the emissivity of the sources along the direction of propagation,
$\omega_{\lambda}$ is the albedo, i.e. the fraction of the opacity
due to scattering, and $\Phi_{\lambda}(cos\,\Theta)$ is the phase
function of the dust grains expressed as a function of $\Theta$, the
angle between the directions determined by the incident radiation
($\theta, \phi$) and the scattered radiation ($\theta^{\prime},
\phi^{\prime}$). One commonly used expression of the scattering phase
function, the Henyey-Greenstein function \citep{hen41}, is characterized by a 
single parameter, the asymmetry parameter $g_{\lambda}$=$<$cos($\Theta$)$>$,
which is the average of the cosine of the scattering angle.

Determining the dust opacity and recovering the intrinsic physical properties
of complex systems like galaxies typically requires the full treatment of the
transfer of radiation through dust (equation~1). The efforts to solve the
integro-differential equation~1 in the case of galactic environments have
generated increasingly sophisticated radiative transfer models and codes to
deal with the complexity of both the dust physical properties and the
dust/emitter geometry. A representative, although non-exhaustive, list of
models and their salient characteristics is given in Table~\ref{tbl-1}. A
detailed description and comparison between these models and codes is beyond
the scope of the present review and the interested reader is referred to the
papers listed in the Table. Although the simplifying assumptions implemented
in many cases often appear to yield as good a solution as more complex
treatments \citep[see, for instance,][]{mis00}, the large number of free
parameters is a fundamental shortcoming of radiative transfer
models. Workarounds to this shortcoming go from assembling large, coherent
sets of independent observational data on just one or a small number of
objects to making over-simplistic assumption on the age and metallicity mix of
the galaxy stellar populations, and often both. This clearly bears deeply on
the generality and applicability of any result and may account to an extent
for the wide variety of conclusions on the opacity of galaxies reached by
different authors.

\subsection{Point Sources and the Interstellar Extinction}

A point-like source (a star or an AGN) behind a screen of dust
is the simplest source/dust geometry, and equation~1 has
solution:
\begin{equation}
I_a(\lambda) = I_o(\lambda) e^{-\tau(\lambda)} = I_o(\lambda) 10^{-0.4 A^o_{\lambda}},
\end{equation}
where I$_a(\lambda)$ and I$_o(\lambda)$ are the attenuated (observed)
and intrinsic radiation intensity, and $\tau(\lambda)=\int
\kappa_{\lambda}\,ds=0.921\,A^o_{\lambda}$ is the optical depth of the
dust screen\footnote{see Appendix~A for the distinction between
A$^o_{\lambda}$ and A$_{\lambda}$ that will be used later.}. The
total-to-selective extinction, i.e. the part of the optical depth
which only depends on the physics of the dust grains and, therefore,
only on the wavelength, is often expressed as
k($\lambda$)=A$^o_{\lambda}$/E(B$-$V). The color excess E(B$-$V) is
defined as A$^o$(B)$-$A$^o$(V), hence k(B)$-$k(V)=1.

Although many studies traditionally employ a `mean Galactic extinction
curve' to correct galactic and extragalactic sources for the effects
of dust obscuration, the well-studied Milky Way interstellar
extinction has been shown to be spatially variable and highly
dependent on the dust grain environment along the line of sight (see
reviews of \citet{mat90} and \citet{fit99}). \citet{ccm89}
demonstrated that the variations can be effectively parametrized by
the total-to-selective extinction at V,
R(V)=A$^o_V$/E(B$-$V). Both the steepness of the far-UV
($\lambda\lesssim$0.3~$\mu$m) rise of the extinction curve and the
strength of the 0.2175~$\mu$m absorption feature decrease for increasing
R(V), as schematically shown in Figure~\ref{fig1} \citep{ccm89}. Most 
sightlines have R(V)$\approx$2--6, with the low values corresponding
to the diffuse ISM (for which R(V)=3.1 is a reasonable average value,
\citet[e.g., ]{rie85}) and the high values pertaining to dense
clouds. The large values of R(V) in dense clouds, usually
accompanied by a gray extinction curve, are consistent with dust grains
having systematically larger sizes than in the diffuse medium
\citep{ccm89}.

The only galaxies external to the Milky Way for which some measurements of the
UV extinction curves exist are the Magellanic Clouds
\citep[Figure~\ref{fig1};][]{koo81,nan81,bou85,fit86,gor98,mis99}, although
tentative measurements in M31 have been made \citep{bia96}. Despite the large
variation of extinction properties found within each of the Magellanic Clouds
\citep[e.g., ]{fit86,leq82,mis99}, the {\em average} curves of the Clouds
appear to have some intrinsic differences from the average Milky Way curve,
including a smaller R(V) (R(V)$\simeq$2.4--2.9, \citet{gor98,mis99}). This has
been interpreted as an effect of the lower metallicity in the Clouds, which
causes the molecular clouds to be more diffuse \citep{pak98} and, possibly, to
produce on average smaller dust grains \citep{mis99}. The weak or absent UV
absorption `bump' at 0.2175~$\mu$m in the extinction curves of the LMC
30~Doradus region and of the SMC bar, respectively, (Figure~\ref{fig1}) could
be related to local processes in these regions, possibly to the ongoing star
formation activity \citep{fit86,gor98,mis99,rea00}. Although the bump carriers
have not been positively identified yet \citep[e.g.,][]{rea00}, mid-- and
far--infrared studies of star forming regions in the Milky Way have provided
evidence that the environmental UV energy density level affects the dust
properties, such as the grain size distribution, via destruction or coagulation,
and/or the dust physical characteristics, via ionization state changes
\citep{boul88,cesa96}. Thus, the environmental UV energy density appears to
play as important a role as the metallicity in shaping the extinction
curve. If confirmed by further observations, this hypothesis may explain the
general absence of a 0.2175~$\mu$m absorption `bump' in the extinction curve
of starburst galaxies (section~4.1) and may partially account for its weakness
in QSOs \citep[e.g.,][]{jura77,spray92,yama99}.

In addition to absorbing stellar light, dust manifests itself through emission
and absorption features and continuum emission at $\lambda>$1~$\mu$m. Features
in the near/mid-IR include, among others, the silicate absorption at 9.7 and
18~$\mu$m, H$_2$O and CO$_2$ ice absorption at 3.0 and 4.3~$\mu$m (see
\citet{lutz96} and \citet{schut98} for a list of other absorption features),
and the emission bands at 3.3, 6.2, 7.7, 8.6, and 11.3~$\mu$m
\citep{sel94,sel00}. The continuum emission beyond a few $\mu$m and up to
$\sim$40--70~$\mu$m is considered to be mainly due to discrete photon heating
of very small dust grains, the smallest among which are transiently heated up
to $\approx$1,000~K \citep[e.g., ][]{sel84,dra85,li01}; dust heated by massive
stars to temperatures $>$70~K also contributes to the flux in the wavelength
region below $\sim$50~$\mu$m, albeit with decreasing filling factor for
increasing temperature \citep{natta76}. Beyond 40--70~$\mu$m, the emission is
mainly due to dust grains in nearly steady balance with the average heating by
starlight; the transition wavelength between predominance of discrete photon
heating versus that of nearly-steady state depends on the average color
temperature of the dust and, thus, on the mean UV energy density level of the
galaxy \citep{mezg82,mat83,helo86,chin86,lons87,row89,des90}.

The original dust model of bare graphite and silicate grains
\citep{mrn77,dra84} has evolved over time by including more complex grain
compositions and extended grain size distributions, to accomodate all the
details of the absorption and emission, and the variability of the
environment, while at the same time accounting for the constraints imposed by the
heavy elements overall abundance \citep[e.g., ]{mat96}. 
Polycyclic aromatic hydrocarbons \citep{leg84,des90,wei00,sel00}, hydrogenated
amorphous carbons \citep{dul89,fur99}, icy and organic refractory mantles
\citep{green84,tiel89,green89,jenn93,green00}, composite and/or porous grains
\citep{mat89,mat96,wol98,vos99}, as well as grain size distributions that
significantly depart from the original power law model \citep{kim94,wei00},
have all been suggested as possible alternatives/complements to the
graphite/silicate model.

\subsection{Extended Emitting Sources}

In the late '80s, \citet{dis89} and \citet{val90} revived the decade-long
debate on whether galaxy disks are mainly transparent or opaque to dust. Using
multi-wavelength information and surface~brightness-versus-inclination tests,
respectively, the two groups of authors reached the similar conclusion that
the disks of galaxies are fairly opaque, with total face-on extinctions in the
B band A$_{B, f}\gtrsim$1~mag\footnote{See Appendix~A for the definition of
A$_{\lambda, f}$.}. These results countered earlier findings, also based on
the surface~brigthness-versus-inclination tests, that galaxy disks are mostly
transparent with face-on extinctions at B less than 1~mag
\citep{holm58,dev59a}.

The reason why similar approaches can produce very different results can be
found in the complexity of the absolute and relative distribution of stars
and dust inside galaxies. When there is a significant departure from the
simplest geometry of a point source (a star) behind a homogeneous screen of
dust, the dust distribution, rather than the optical depth of the dust layer,
becomes the dominant factor in determining the obscuration suffered by the
stellar radiation \citep{natta84,wit92,cal94}. In particular, differential
optical depth effects come into play. If, for instance, the stars are
homogeneously mixed with the dust, the shorter the wavelength of the radiation
the smaller the optical depth of the dust layer from which it can emerge;
thus, an observer will detect blue light only from the closest outer `skin' of
the mixed distribution, while redder light will come from deeper layers. An
extreme case is that of infinitely optically thick clumps in front of an
extended light distribution: the stellar emission will come only from those
lines of sight that do not intersect clumps, and the net result will be an
unreddened, albeit dimmer than the original, spectral energy distribution
(SED). The scattering component of the extinction curve will also modify the
general characteristics of the dust obscuration, as light will be scattered
{\em into the line of sight} as well as out of it. The net result of all these
effects is a general blueing of the emerging SED relative to the foreground
homogeneous dust screen case, or, in other words, an often much greyer
`effective extinction', even in the presence of large dust optical
depths\footnote{See Appendix A for the relation between optical depth,
extinction, and effective extinction.} \citep{bru88,wit92}.

For illustrative purposes, Figures~\ref{fig2} and \ref{fig3} show how
photometric and spectroscopic quantities are affected by the dust geometry for
five `toy models'.  The models represent plane-parallel geometries of dust and
stars \citep[details are given in][]{natta84,cal94,wan96} for: (1) the simple
case of a homogeneous, non-scattering dust screen foreground to the light
source; (2) a homogeneous, scattering dust screen foreground to the light
source (e.g., a screen in close proximity of the source, so that scattering
into the line of sight becomes an important effect); (3) a Poissonian
distribution of clumps in front of the light source, with average number $\cal
N$=10; (4) a homogeneous mixture of dust and emitters; (5) a homogeneous
dust/emitter distribution averaged over all inclination angles, with the UV
and ionized gas emission having the same scaleheight of the dust while the
scaleheight of the optical light gradually increases to twice that of the dust
at V and beyond. Models 2--5 include scattering in the
calculations\footnote{For the models, the compilation of albedo and asymmetry
parameter values of \citet{wit00} is used. In this compilation, the data are
mainly from studies of reflection nebulae, and the 0.2175~$\mu$m feature is
due to pure absorption (no scattering) \citep{cal95a}. These $a$ and $g$
values are assumed here to be representative of dust properties, although the
following caveats need to be kept in mind: some values, especially for the
albedo, are still unsettled \citep{burgh01}; the generalization of the
reflection nebulae results (R(V)$>$3.1) to the diffuse dust is not immediate
\citep{wit97,kim94,wei00}; the UV $a$ and $g$ values could be lower in the
Magellanic Clouds than in the Milky Way \citep{wei00}.}.  In all cases, the
light source is assumed to be extended and to be represented by a 300~Myr old
stellar population undergoing constant star formation, with solar metallicity
and a Salpeter IMF in the mass range 0.1--100~M$_{\odot}$ \citep{lei99}.

The `blueing' effect of geometries more complex than a foreground dust screen
is evident from the comparison of the five models. At fixed optical depth
$\tau_V$ (fixed amount of dust), the reddening of the V$-$I color of a stellar
population induced by a generic geometry of stars and dust will be less, and
the effective extinction A$_V$ lower, than that induced by a foreground screen
(Figure~\ref{fig2}, panels~a and b). The same will be true for colors at
other wavelengths (Figure~\ref{fig2}), where, again, the foreground
non-scattering screen will provide the maximum reddening. For mixtures of dust
and stars (models~4 and 5), reddening of colors and effective extinction reach
asymptotic values; beyond certain values of the optical depth, an increase in
$\tau_V$ corresponds to small or negligible changes in colors and A$_V$.  This
is due to the fact that the shortest wavelength emission is contributed only 
by the dust layers closest to the observer.  One common consequence of the
greyer 'effective extinction' caused by complex dust/emitter geometries is
that a limited number of diagnostics over a small wavelength range will lead
to underestimates of the true galaxy opacity, especially if the actual dust
geometry is not known and is `a priori' assumed to be foreground
\citep{wit92}.

To aid the comparison with data on local starbursts and high-redshift
star-foming galaxies (sections~4 and 5), the models shown in
Figure~\ref{fig3} are a function of the ratio of the
attenuated-to-intrinsic Balmer line ratio:
\begin{equation}
R_{\alpha\beta}={(L_{H\alpha}/L_{H\beta})_{a}\over
(L_{H\alpha}/L_{H\beta})_{o}},
\end{equation}
and of the UV spectral slope\footnote{see Appendix~B for the definition of
$\beta$.} $\beta$.  The luminosity ratios of nebular hydrogen lines, like
H$\beta$, H$\alpha$, Pa$\beta$, Br$\gamma$, etc., are rather insensitive to
the details of the underlying stellar population and of the IMF, being
affected at the $\sim$5\%--10\% level by variations of the gas physical
conditions \citep[e.g.,][]{oste89}; they represent, therefore, accurate probes
of dust reddening and geometry. Table~\ref{tbl-2} lists the commonly used
hydrogen emission line ratios, and the differential extinction between each
pair. When multiple line ratios widely separated in wavelength are
available, they can constrain the dust geometry in the sampled wavelength
range. The two drawbacks of using nebular lines are: (1) they are produced by
ionizing photons, therefore an ionizing stellar population needs to be present
in the galaxy; and (2) they only probe the attenuation of the {\em ionized
gas}, and such attenuation may or may not be directly related to that of the
stellar population (section~4). The explicit expression of R$_{\alpha\beta}$
depends on the dust geometry; in the simple case of a foreground,
non-scattering screen:
\begin{equation}
R_{\alpha\beta}=10^{0.4 E(B-V) [k(H\beta)-k(H\alpha)]}, 
\end{equation}
where the values of k(H$\alpha$) and k(H$\beta)$ are given in
Table~\ref{tbl-2}.

Degeneracies between the reddening induced by dust and that induced by
variations in the age, metallicity, and IMF of the stellar populations play an
non-negligible role in opacity determinations of galaxies. These effects are
especially important when only broad band colors or low resolution
spectroscopy are available, and age-sensitive or IMF-sensitive stellar
features are not identifiable \citep[e.g., discussion in][]{pap01}. For the
age-dust degeneracy, an example is given in Figure~\ref{fig4}; here, the broad
band colors and magnitudes of a dust-free, ageing stellar population produced
by an instantaneous burst of star formation are compared with the analogous
quantities of an increasingly extincted 6~Myr old stellar population. A clumpy
dust distribution with $\tau_V\sim$2 in front of the young population reddens
and attenuates it enough that its broad-band UV, optical, and near-IR colors
closely resemble those of a 20~Myr old, dust-free population (corresponding to
the color V$-$I$\sim$0.6 in Figure~\ref{fig4}). Thus, the dusty, young
population can be mistaken for a dust-free population a factor $\sim$3
older. This example is just one out of a large range of possibilities in the
multi-dimensional dust/age/IMF/metallicity parameter space.

In summary, the grey effective extinction produced by dust geometry and
scattering, the age-IMF-metallicity-dust degeneracy, together with
uncertainties on the appropriate extinction curve to be applied to different
galaxies (section~2.1) and the absence of strong emission and
absorption features below $\sim$1~$\mu$m (apart from the $\sim$0.22~\AA~
absorption bump which is not universal), all contribute to make dust
obscuration measurements very difficult in galaxies.

\section{Measurements of Dust Opacity in Local Galaxies}

\subsection{Spiral Disks}

Efforts to determine the dust opacity of local galaxies have spawned
the development of a number of independent techniques that have been
applied to a large array of samples. The most widely used methods in
the case of galaxy disks are tests of the dependence of the surface
brightness on inclination, multi-wavelength comparisons, and
statistical analysis of the color and number count variations induced
by a foreground galaxy onto background sources. Each method has its
strengths and weaknesses, as briefly detailed below.

As \citet{bur91}, \citet{cho91},and \citet{dav93} pointed out, probing galaxy
dust opacity by measuring the surface brightness dependence of galaxy disks on
inclination can suffer from drastic selection biases. For increasing
inclination, an apparent-magnitude-limited catalog will mimick the condition
for transparent galaxies, with mildly increasing surface brightnesses,
increasing diameters, and constant total luminosities. Conversely, an
apparent-diameter-limited catalog will mimick the conditions for opaque disks,
with roughly constant surface brightnesses, increasing magnitudes and constant
diameters as a function of increasing inclination \citep{bur91}. Selection
effects can, however, be controlled by combining different samples and by
using galaxy redshifts/distances to discriminate among the different biases,
and analysis of large samples of late-type (mainly Sbc and Sc) spirals
\citep{hui92,pel92,gio94,gio95,jon96,mor98} has shown that the central regions
of disk galaxies are opaque while the external regions beyond 2--3 scale
lengths are almost completely transparent; there is also a trend for luminous
galaxies to be more opaque than less luminous galaxies \citep{gio95}. In
bright, M$_I<-$21 disk galaxies, the observed magnitude correction from
edge-on to face-on, $\Delta$m$_I\sim$1~mag, corresponds to a face-on central
optical depth\footnote{The face-on central optical depth $\tau(0)$ is the
optical depth of the dust at the geometrical center of the tri-dimensional
dust distribution for a face-on disk galaxy.} $\tau_V$(0)$\gtrsim$5; this
implies effective face-on extinctions A$_{B, f}\sim$0.4--0.5~mag, A$_{I,
f}\sim$0.2~mag, and A$_{H, f}\sim$0.1~mag \citep{pel92}, for a double
exponential distribution of dust and stars, and for practically transparent
external galaxy regions \citep{byu94}. Less luminous galaxies are
progressively less opaque in their central regions, with a edge-on~to~face-on
correction of $\Delta$m$_I\sim$0.5~mag at M$_I>-$18, corresponding to
effective face-on extinctions A$_{B,f}<$0.1~mag \citep{gio95}.

Qualitative results similar to those above have been obtained by \citet{bos92}
and \citet{byu93} for a small number of edge-on late-type spirals using
multi-wavelength kinematic information. The method, proposed by \citet{goa81},
consists of comparing rotation curves of an edge-on galaxy at widely different
wavelengths. If the edge-on galaxy is optically thick at a given wavelength,
say, H$\alpha$, the rotation curve measured at this wavelength shows apparent
solid body rotation throughout the entire disk, because the H$\alpha$ emission
is coming only from the external regions of the galaxy. A comparison with a
mostly unextincted rotation curve measured at a longer wavelength, such as HI,
can then establish whether optical thickness at H$\alpha$ exists. This method,
though fairly direct in principle, may suffer from two limitations: first,
dust in galaxies tends to be patchy and the H$\alpha$ emission may be detected
along the least extincted sight lines even in the presence of considerable
opacity; second, the method implicitly assumes that H$\alpha$-emitting regions
are uniformly distributed across the disk.

The multi-wavelength comparison method provides an independent approach to the
problem of determining the dust opacity of galaxies. In broad terms, the
technique consists of using multiwavelength data to solve simultaneously for
the intrinsic colors of the stellar populations and for the
reddening/attenuation induced by the dust. Generally, broad band images
covering as large as possible wavelength baseline, often from B to K band, are
used to determine the opacity of the disks, by exploiting the differential
extinction property of dust
\citep{rix93,blo94,pel95,ems95,bec96,kuc98,xil99}. Indeed, in face-on
quiescent galaxies K-band emission appears negligibly affected by dust
obscuration, with $\lesssim$10\% of the light lost to dust absorption and
scattering \citep{rix93}.  The multiwavelength technique is the most widely
adopted for measuring dust opacity in galaxies, because it employes easily
accessible information on diverse galaxy types, and enables detailed modelling
of optical depths as a function of galactic region. There are, however, three
main weaknesses to the method: first, this technique needs as wide a
wavelength coverage as possible, often difficult to obtain, to yield
unambiguous results; second, if the dust optical depth exceeds unity at some
wavelength, different volumes of the galaxy will be sampled at different
wavelengths; third, the need for large datasets and detailed modelling makes
obtaining results on large samples relatively cumbersome. Results vary from
author to author, due to the different sample selections, galaxy and dust
distribution models, and measurement techniques applied. Optical and near-IR
image analysis gives typical central optical depths $\tau_V$(0) in the range
0.3--2.5 for types Sab-Sc \citep{pel95,kuc98,xil99}, although higher values
have been obtained by \citet{blo94}. For exponential stellar and dust
profiles, these values of $\tau_V$(0) result in optical depths
$<\tau_V>\lesssim$1 averaged over the inner 2--3 scale-lengths
\citep{rix93,bec96,kuc98}, and A$_{B, f}\sim$0.3--0.4~mag and A$_{I,
f}\lesssim$0.15~mag over the same area. Much lower values of the optical depth
are found for late-type, low surface brightness galaxies, with face-on,
central values $\tau_V$(0)$\sim$0.15 \citep{mat00}. In galaxy disks, arm
regions tend to be opaque, with $\tau_V\sim$1.4--5.5 (A$_{I,
f}\sim$0.2--0.7~mag) and peak values at $\tau_V\sim$10--12, while interarm
regions are generally transparent, with $\tau_V\sim$0.3--1.1 (A$_{I,
f}\sim$0.05--0.15~mag, \citet{elm80,rix93,bec96}, but see the higher interarm
extinction values found by \citet{tre98}).

Energy balance estimates between dust absorption and emission can also be
classified as a `multiwavelength technique', but it is conceptually different
from the one above. This technique employes direct measurements of the dust
emission in the infrared to derive an estimate of the total stellar energy
absorbed by dust, and then derive opacities at shorter wavelengths; whenever
possible, the wavelength baseline is extended to the UV to probe the direct
light from massive stellar populations and constrain bolometric quantities
\citep{xu95,xu96,bua96,wan96,tre97,tre98}. By combining UV and infrared
(IRAS) data, \citet{xu95}, \citet{wan96}, and \citet{bua96} find typically
lower $<\tau_V>$ values than those derived from optical/near-IR images. This
is not incompatible with those results, since \citet{xu95} and
\citet{wan96}'s models assume plane-parallel dust and stars, instead of
exponential profiles, and thus the opaque inner regions are averaged with the
nearly transparent outer regions \citep{pel95,kuc98}. Still, the UV
selection of the samples may bias the conclusions towards lower optical depths
and the use of IRAS data leaves some uncertainty as to the bolometric
correction of the dust emission. The use of ISO data, that extend the
infrared wavelength coverage out to 200~$\mu$m, appear indeed to favor
higher values of the effective extinction \citep{tre98}. \citet{wan96} find
that opacity in disk galaxies is a function of luminosity
(Figure~\ref{fig5}), in agreeement with \citet{gio95}. These authors
parametrize the opacity as: $<\tau_V>$=$<\tau_{V,*}>(L_V/L_{V,*})^{0.5}$, with
$<\tau_{V,*}>$=0.6 and L$_{V,*}$ the visible luminosity of a L$^*$ galaxy in
the local Universe. Similarly, \citet{bua96} note that Sb--Scd galaxies have
$<\tau_V>\sim$0.8 and are on average 1.5 times more opaque than Sa--Sab, and
$\sim$3 times more opaque than Sd--Irr galaxies, with effective face-on
extinctions A$_{B, f}\sim$0.3~mag, A$_{B, f}\sim$0.15~mag, and A$_{B,
f}<$0.1~mag for Sb--Scd, Sa--Sab, and Sd-Irr, respectively.

All methods described so far use the stellar radiation of the galaxy itself as
light source for the dust, with the complication that intrinsic colors, age
and metallicity gradients of the stellar populations and the relative
distributions of dust and stars need to be disentangled from the extinction
proper. A technique that overcomes these problems is the one that uses
background sources as `light bulbs' for the dust in foreground galaxies. In
this case, the dust opacity is measured across the full thickness of the
foreground galaxy. There are basically two approaches to the technique. In the
first approach the foreground galaxy is partially projected onto a nearby
background galaxy \citep{kee83,whi92,whi00,whi01,kee01,ber97,piz98}. In
general, the background galaxy needs to present a smooth light profile, thus
early types and preferably ellipticals are required \citep[see,
however,][]{kee01}; the dimming and color changes induced by the foreground
galaxy dust onto the isophotes of the background source provides a measurement
of dust opacity. The drawbacks of this approach are twofold: (1) there are
very few known pairs of galaxies that fullfil all the requirements and (2) the
method is limited by the degree of symmetry of both galaxies
\citep{whi00}. The second approach uses distant galaxies as background light
sources. Statistical variations in number counts and colors of the background
galaxies relative to control fields provide a handle on the total dust optical
depth of the foreground galaxy \citep{gon98}. This method overcomes most of
the limitations of the first one, but it presents some of its own: (1) high
angular resolution images (e.g., from HST) are needed in order to identify
background galaxies; (2) even with such images it is often difficult to
discriminate background sources from extended regions within the foreground
galaxy (e.g., HII regions), especially in crowded areas. Because of these
limitations, the second method may also tend to avoid crowded and/or heavily
extincted areas, thus biasing the results towards less dusty galactic
regions. Results from various authors indicate that the spiral arms of
galaxies tend to be opaque with A$^o_B\sim$0.5--2~mag and
A$^o_I\sim$0.3--1.5~mag at any radius, while the opacity of interarm regions
decreases with increasing distance from the galaxy center, starting at
A$^o_I\approx$0.7~mag within 0.3~R$_{25}$ and decreasing to practically zero
at R$_{25}$ \citep{gon98,whi00,kee01}. The dust masses inferred from these
extinction measures agree within a factor $\sim$2 with masses derived from the
infrared/submm (ISO and SCUBA) dust emission \citep{dom99}. One common
trend generally found is the greyer-than-Galactic reddening in the arm regions
\citep[see, however,][]{kee01}. The clumpiness of the dust within each
resolution bin has been suggested to be responsible for the grey reddening
\citep{ber97,gon98,whi00,whi01}. Less agreement exists on the shape of the
reddening curve in the interarm regions, due to the larger uncertainties
resulting from the lower extinction values. The measured extinctions are
front-to-back, i.e. represent the optical depth of the intervening dust
between the observer and the background sources. To relate these numbers to
those derived from the inclination and multi-wavelength tests, one has to make
assumptions as to the distribution of stars relative to the dust in the
foreground galaxy; for a mixed geometry of stars and small dusty clumps, the
measured optical depths correspond to effective I extinctions of A$_{I,
f}\sim$0.3--0.4~mag and A$_{I, f}\lesssim$0.2~mag for the arms and the
interarm regions, respectively.

\subsection{Beyond the Optical Disks}

Multi-wavelength optical and near-IR imaging of edge-on or nearly edge-on
systems indicate that the scalelength of the dust is about 40\% larger than
the scalelength of the stars \citep{xil99}, while the dust/stars scaleheight
ratio is in the range 0.25--0.75, with a mean value of 0.5,
\citep{kyl87,xil97,xil99} for late-type spirals (Sb and later) and appears to
have ratio $\sim$1 in early-type spirals \citep{wai90}. IRAS 100~$\mu$m
\citep{nel98} and ISO long wavelength maps \citep{alt98,dav99,tre00,rado01}
confirm that cold dust emission extends beyond the limits of the optical disks
along the radial direction, with scales that are $\sim$40\% larger than those
of the B-band emitting stars, but still well within the HI disks
\citep[e.g.,][]{martin98}. Using statistical measurements of color variations
between background galaxies and control fields, \citet{zar94} give a
$\gtrsim$2~$\sigma$ detection of A$^o_I\approx$0.04 in galaxy haloes, at a
distance of about 60~kpc from the galaxy centers, along the optical major
axis.

\subsection{Irregular Galaxies}

Spiral and irregular galaxies follow a global metallicity-luminosity (and
metallicity-mass) relation, where the oxygen abundance increases by a factor
$\sim$100 for an increase of $\sim$10$^4$ in absolute blue luminosity
\citep{ski89,zar94b}. Irregulars occupy the low metallicity locus of the
relation, with $\approx$10 times on average lower oxygen abundance than
spirals. Since the HI masses per unit L$_B$ in irregulars are a factor
$\approx$3 larger than in spirals \citep{rob94,zar94b}, the mass in
interstellar metals per unit L$_B$ is $\approx$3 times smaller. Adding the
H$_2$ gas into the balance does not change this conclusion, since the H$_2$/HI
mass ratio in irregulars appear to be no larger than that in spirals, even
accounting for the uncertainties in the CO-to-H$_2$ conversion factor
\citep{verter95,arim96,wils97}. Measurements give the H$_2$ mass as 25\%--50\%
of the HI mass in late spirals and as $\approx$20\% of the HI mass in
irregulars \citep{you91,hun93,hun96,hun97,mei01}. Smaller metal masses for the
irregulars translate into smaller dust masses per unit L$_B$, for standard
values of the metal-to-dust ratio \citep{van74,leq84}, and into the
expectation that irregular galaxies are at least 3 times more transparent at B
than spiral galaxies, for similar dust/star geometries.

The analysis of the blue, H$\alpha$, and IRAS-detected infrared emission
confirms that irregular galaxies are relatively transparent systems
\citep{hun86,hun89}. In a study of the Magellanic Clouds using background
galaxies, \citet{dutr01} find that the central regions have A$^o_B\gtrsim$0.25
in the LMC and A$^o_B\gtrsim$0.16 in the SMC. In their UV selected samples,
\citet{wan96} find that a 0.1~L$^*$ galaxy has a mean optical depth about 1/3
of the optical depth of an L$^*$ galaxy, and \citet{bua96} reach a similar
conclusion by comparing Sb-Scd galaxies with Sd-Irr. Thus, if the typical
face-on effective extinction for an Sc galaxy is A$_{B, f}\sim$0.4--0.5~mag,
the analogous quantities for an irregular are A$_{B, f}\sim$0.15 and A$_{I,
f}\sim$0.05--0.10.

\subsection{Early Hubble Types}

Although early Hubble type galaxies (ellipticals and S0s) do not
generally fall under the category of `star-forming' galaxies, they
occupy a niche in terms of dust opacity properties that is at the same
time opposite and complementary to that of other galaxies. Ellipticals
and S0s are at the opposite end of the Hubble classification relative
to Irregulars, but they appear to be no more opaque than those late
systems.

Early type galaxies cover a metallicity range similar to that of late type
spirals \citep{bro91,zar94b}, but have total (HI$+$H$_2$) gas fractions per
unit optical luminosity that are $\sim$3 times or more smaller
\citep{you91,rob94}. Thus, for standard interstellar metal-to-dust ratios,
early type galaxies are expected to be more transparent than disk galaxies.

The IRAS survey has shown that between 10\% and 50\% of all nearby early type
systems (45\% of ellipticals and 68\% of S0s) contain dust
\citep{jur87,kna89,bre98}, although the involved masses are generally modest,
typically in the range 10$^5$--10$^6$~M$_{\odot}$ or
M$_{dust}$/L$_B\approx$10$^{-5}$--10$^{-4}$~M$_{\odot}$~L$_{\odot}^{-1}$ 
\citep{rob91,goud95,wis96,bre98}. The exact fraction of dusty systems  
appears to depend upon the adopted detection threshold \citep{bre98}, but it
is likely a lower bound to the actual value, given the limits in the IRAS
sensitivity. The 50\%~ figure seems in agreement with the results
of optical surveys: at least 40\% of the local ellipticals and S0s show
presence of disks, lanes and/or patchy distributions of dust in visual images
\citep{ebn85,sad85,ebn88,goud94}. About 40--50\%, and possibly as many as
80\%, of ellipticals also show evidence for circumnuclear dust on scales
$<$100~pc, as seen in HST images; the small-scale dust is mostly associated
with nuclear radio activity \citep{van95,lau95,for95,car97,tra01}.

Despite the abundance of evidence for dust, optical extinction determinations
of dust features in ellipticals generally account for only $\sim$1/10 to
$\sim$1/5 of the dust mass measured from the infrared (IRAS) emission; the
remaining 80\%--90\% of the dust mass is probably associated with a component
smoothly distributed across the galaxy, and thus difficult to detect with
standard techniques \citep{goud95,wis96,bre98,mer98}. Limited data from ISO in
the wavelength range 120--200~$\mu$m confirm these findings
\citep{haa98}. Using the elliptical galaxy model of \citet{wit92},
\citet{goud95} determine that the observed infrared luminosities are
compatible with central optical depths of the diffuse component 
$\tau_V$(0)$\lesssim$0.7, with a typical value $\tau_V$(0)$\sim$0.2--0.3 (see,
however, the somewhat larger values predicted by the models of
\citet{wis96}). The corresponding effective extinction is
A$_B\lesssim$0.05--0.10~mag and A$_I\lesssim$0.02--0.04~mag.

The giant ellipticals found at the centers of cooling flow clusters are often
surrounded by extended, $\approx$10--100~kpc, filamentary, and dusty
emission-line nebulae, with a higher frequency in those clusters with cooling
times shorter than a Hubble time \citep{heck89,spa89,hu92,voi97,don00}. Dust
lanes in the central galaxies have, indeed, been directly observed
\citep{pin96}. Optical and UV emission line studies give extinction values in
the range A$_V\sim$0.3--2~mag for the dust associated with the nebula
\citep{spa89,hu92,all95,voi97,don00}; infrared and sub--mm measurements
give dust masses of the order of 10$^7$--10$^8$~M$_{\odot}$, or
M$_{dust}$/L$_B\approx$10$^{-4}$--10$^{-3}$~M$_{\odot}$~L$_{\odot}^{-1}$, for
these systems \citep{cox95,edg99}. The existing controversy on the origin of
the dust associated with the central galaxies of cooling flow clusters will be
briefly reviewed in section~6. Such galaxies are, however, not a statistically
significant component of ellipticals in general, thus their contribution to
the dust mass budget of early type galaxies is small.

\subsection{Infrared-Bright Galaxies}

The galaxies discussed so far tend to be mostly `quiescent', i.e. their
bolometric output is {\em not dominated} by a single region of active 
star formation or by nuclear non-thermal activity, and have 
modest infrared outputs. Their infrared luminosities in the
3--1000~$\mu$m band are typically $\approx$10$^9$~L$_{\odot}$ and less
than $\sim$4--6$\times$10$^{10}$~L$_{\odot}$, with 
infrared-to-blue luminosity ratios L$_{IR}$/L$_B\sim$0.4--0.5
\citep{dej84,soi87,dev91,soi91,hun96}, and $\approx$1/2 or less of
the stellar bolometric luminosity converted into infrared emission
via dust reprocessing\footnote{The total infrared output L$_{IR}$ of
the dust emission is derived from the infrared luminosity measured in
the 8--120~$\mu$m or 40--120~$\mu$m IRAS window \citep{hel88,san96}
multiplied by $\sim$1.4 or $\sim$2.2, respectively, to account for
bolometric correction factors as modeled by \citet{dal01} from ISO
data. The blue luminosity is L$_B\propto\lambda
f(\lambda)$(0.44~$\mu$m).} (see Table~\ref{tbl-3}).

Galaxies with L$_{IR}>$6$\times$10$^{10}$~L$_{\odot}$ and
L$_{IR}$/L$_B>$1 are progressively associated with dustier and more
active systems for increasing values of the infrared luminosity
\citep{rie86}. The brightest infrared sources are not necessarily
associated with the most massive galaxies, despite the existence of a
relation between a galaxy luminosity (mass) and its dust content
\citep{wan96}. Infrared luminosity is a combination of dust content
and activity: brighter sources have larger amounts 
of star formation and/or non-thermal activity embedded in dust. Most
of the luminosity from galaxies with
L$_{IR}\approx$L$_{bol}\gtrsim$10$^{11}$~L$_{\odot}\sim$3--4~L$^*$
appears associated with sources heavily obscured by dust
\citep{san96}. At the bright end of the infrared luminosity function
\citep{soi87,soi91,kim98} are the Ultraluminous Infrared Galaxies
(ULIRGs) with luminosities from L$_{IR}\sim$10$^{12}$~L$_{\odot}$ up to
$\sim$8$\times$10$^{12}$~L$_{\odot}$ \citep{soi87,kim98},
infrared-to-blue ratios around 30--400 \citep{san96,cle96}, and
optical extinctions in excess of 10--40 magnitudes
\citep{gen98,mur01}.  These sources show a continumm of properties as
a function of increasing luminosity, with the fraction of
starburst-dominated ULIRGs going from $\sim$80\%--85\% at the low
luminosity end to $\sim$50\% for
L$_{IR}>$2$\times$10$^{12}$~L$_{\odot}$, and the remaining fraction
being AGN-dominated \citep{gen98,lut98,vei99}. Comprehensive reviews
of the properties of ULIRGs are given in \citet{san96} and \citet{gen00}.
 
Despite their large luminosities, bright infrared sources do not
represent a major contributor to the energy budget of galaxies in the
present-day Universe. Such objects are relatively rare within the
local 100--200~Mpc: galaxies with L$_{IR}\ge$10$^{11}$~L$_{\odot}$ and
with L$_{IR}\ge$10$^{12}$~L$_{\odot}$ represent $\sim$6\% and $<$1\%,
respectively, of the total infrared emission in the local Universe
\citep{soi87,san96,kim98,yun01}. However, the contribution of
infrared--bright sources to the galaxy energy budget increases with
redshift (see section~5.2), as does their relevance in the framework
of galaxy evolution.

\subsection{Summary}

Although the issue of dust opacity of galaxies is far from settled, an overall
picture is apparent: galaxies in the local Universe are only moderately
opaque, and extreme values of the opacity are only found in the statistically
non-dominant, more active systems.

Table~\ref{tbl-3} summarizes the main conclusions of this section, by listing
the effective extinction values for local galaxies according to morphological
type. For disk and irregular galaxies, both the face-on and the
inclination-averaged extinctions are given. For instance, in Sb-Scd galaxies,
the typical inclination-averaged extinctions are 0.6--0.65~mag larger than
face-on at 0.15~$\mu$m, $\sim$0.4~mag at B, $\sim$0.25~mag at I, and 0.1~mag
at K, for a stellar to dust scaleheight ratio that varies between 1 in the UV
and 2 in the I and K bands (model~5 of section~2.2).

Among the non-active galaxies, luminous Sb--Scd galaxies contain the
largest amounts of dust, which absorb $\approx$1/2 of their bolometric
light. For comparison, dust emission represents between 35\% and 40\%
of the bolometric luminosity of our Milky Way galaxy
\citep{cox86,sod94}. The dust content generally decreases in less
luminous galaxies and for earlier and later type galaxies; dust
absorbs less than $\sim$15\% of the bolometric light in E/S0 galaxies,
and the fraction rises to $\approx$30\%~ in irregulars
(Table~\ref{tbl-3}).

Interpolating from Table~\ref{tbl-3}, the inclination-averaged effective
extinction of the stellar emission in late spiral galaxies at the restframe
wavelength of H$\alpha$ is A$_{0.6563}\lesssim$0.60~mag. This value should be
compared with the effective extinction measured for the H$\alpha$ line
emission, A$_{H\alpha}\sim$0.8--1.1~mag, from the ratio of the line to the
reddening-free thermal radio emission
\citep{ken83,nik97,ken98,bel01}. Althought there are relatively large
uncertainties associated with both numbers, the two extinction values suggest
that the nebular emission is more reddened than the stellar continuum, with
A$_{H\alpha}\approx$1.5~A$_{0.6563}$. Studies of controlled samples indicate a
value of 2 as more typical for the ratio between the dust attenuation of the
ionized gas and that of the stellar continuum, both in the case of spirals and
irregulars \citep{bel01b} and in the case of moderate-intensity starbursts
(section~4).

\section{A Case Study in the Local Universe: Starburst Galaxies}

Among local galaxies, starbursts occupy a special niche.  The average
UV and infrared luminosities (both proxies for SFR) per unit comoving
volume increase with redshift by roughly an order of magnitude up to
z$\sim$1--2 \citep{lil96,cow99}. Active star formation played a larger
role in the past than today, and may have been a fundamental
ingredient in the shaping of galaxy populations.

Nuclear and circumnuclear starbursts occur in galaxies spanning a wide
range of properties, from almost dust-free, infrared-faint, low-mass
irregulars (e.g., Tol1924$-$416 and NGC1569) to dust-rich, massive
spirals (e.g, NGC5236), to infrared-bright interacting galaxies
(e.g., NGC7714 and NGC6090), to mergers and ULIRGs \citep[e.g.,
NGC4194 and NGC1614; Arp220,][]{sco98}. Nuclear and circumnuclear
bursts of star formation can be triggered by a variety of mechanisms,
including secular evolution of bars \citep{fri95,nor96} and galaxy
interactions and mergers \citep{lar78,kee85,san88,mih96,hib97}. The
latter produce the most luminous starbursts (ULIRGs), which
rapidly deplete the molecular gas content of the galaxy as merging
progresses \citep{yu99}. Less extreme starbursts, hosted in galaxies
with L$_{bol}\lesssim$5$\times$10$^{11}$~L$_{\odot}$, obey a positive
correlation between the luminosity of the starburst and the luminosity
and mass of the host galaxy \citep{heck98}:
\begin{equation}
L_{B, host} (L_{\odot})\sim 10^{9.9} SFR_{starb}^{0.94}, 
\end{equation}
where SFR$_{starb}$ is in M$_{\odot}$~yr$^{-1}$ for a Salpeter IMF in the mass
range 0.1--100~M$_{\odot}$ \citep[from the formulae of][]{ken98}. Equation~5
is probably a consequence of the fact that large SFRs are substained by the
massive gas inflows that only a deep potential well can support
\citep{heck98}. Indeed, from causality arguments, the maximum luminosity of a
starburst scales as $\approx\sigma$v$^2$, with $\sigma$ the dispersion
velocity of the gas and v the rotation velocity of the host galaxy
\citep{heck94,leh96,meu97}. There are notable exceptions to this average
trend: M82 is an example of a dwarf, M$\sim$10$^{10}$~M$_{\odot}$
\citep{sof92} and L$_{B, host}\sim$5$\times$10$^9$~L$_{\odot}$, galaxy hosting
a relatively powerful starburst, with SFR$\sim$6~M$_{\odot}$~yr$^{-1}$ and
L$_{IR}\sim$3$\times$10$^{10}$~L$_{\odot}$. A secondary effect of both the
SFR$_{starb}$--host~galaxy~mass correlation and the mass-metallicity
correlation is that the more powerful starbursts are hosted in the more
metal-rich and dustier host galaxies \citep[][also Figure~\ref{fig7},
panel~c]{heck98}:
\begin{equation}
10^{A_{0.15}}\sim 250 SFR_{starb}^{2.2}
\end{equation}
\citep[see, also,][]{hopk01}. Although equation~6 is not applicable to
ULIRGs, still in the local Universe there are no examples of `naked'
Arp220-like galaxies, emitting most of the luminosity produced by
their SFR$\sim$100--200~M$_{\odot}$~yr$^{-1}$ directly in the
UV. Objects like Arp220 tend to be heavily extincted, with
A$_V\sim$15--45~mag for Arp220 itself \citep{sco98,gen98,shi01}, and
to be bright infrared sources (section~3.5).

The rest of section~4 discusses the dust reddening and obscuration
properties of the local starburst galaxies with
L$_{bol}\lesssim$5$\times$10$^{11}$~L$_{\odot}$. The specific sample
contains starbursts that are UV--luminous enough to have been detected
by IUE \citep{kinn93}; it will be termed `UV-selected' in what
follows.  The sample is not statistically complete or rigorously
defined; however, it is representative of the broad range of
properties of local UV--bright starbursts \citep{heck98}. These
galaxies are not as extreme in dust extinction and activity level as
ULIRGs. They have SFRs per unit area in the range
$\sim$0.3--20~M$_{\odot}$~yr$^{-1}$~kpc$^{-2}$ and cover a limited
range in the parameter space of optical attenuation,
A$_{H\alpha}\sim$0--2.5~mag. The attenuation refers to the integrated
value across the starburst site; individual regions or clumps can have
effective extinctions as high as A$_{H\alpha}\sim$30~mag
\citep{beck96,cal97b}. The host galaxies cover a wide variety of late
type morphologies, from grand design spirals to irregular and
amorphous, to distuberd morphologies that result from interactions or
merging processes. The starbursts mostly occupy the inner region of
solid body rotation of the host galaxy, with sizes $\sim$0.5--4~kpc
\citep{leh96}. The UV selection \citep{kinn93,cal94} aids the
discrimination from their more opaque and infrared-brighter
counterparts. Although the UV selection results in some limitations to
the applicability of the obscuration properties discussed below (see
end of section~4.2), it does not entirely exclude dusty systems; the
sample contains objects in which up to 90\%--95\% of the bolometric
energy is re-processed by dust. Table~\ref{tbl-4} gives three
representative examples each for both the high-luminosity, dust-rich
end and the low-luminosity, dust-poor end of the UV-selected
starbursts, together with the average of the sample.

\subsection{Dust Obscuration Characteristics of Starbursts}

Although the UV-selected starbursts form an heterogeneous set, their dust
reddening and obcuration properties are remarkably uniform. Higher dust
optical depths produce redder UV--to--near--IR SEDs and nebular
emission line ratios (Figure~\ref{fig6}), almost independent of the details
of the dust distribution and of the intrinsic stellar population 
\citep{cal94,cal97}.

The geometry of the dust that best describes the reddening of the
ionized gas emission in these systems, in the wavelength range
0.48~$\mu$m--2.2~$\mu$m, is that of a foreground--like
distribution. Figure~\ref{fig6}, panel~b, shows the comparison
between observations and the models of section~2.2 for the
attenuated-to-intrinsic hydrogen line ratios H$\alpha$/H$\beta$ and
H$\beta$/Br$\gamma$. Internal dust does not appear to be a major
component in the starburst region(s), and the little present is likely
to be in compact clumps \citep{cal95b}. The main source of opacity
appears to be given by dust that is external, or mostly external, to
the starburst region (although still internal to the host galaxy),
similar to a (clumpy) dust shell surrounding a central starburst. This
is, of course, verified only in the B--to--K wavelength range where
the reddening of the nebular emission lines has been measured;
measurements at longer wavelengths could easily reveal more complex
geometries. From a practical point of view, the data in
Figure~\ref{fig6}, panel~b, are well described by a simple
foreground, non-scattering dust distribution, with MW extinction curve
for the diffuse ISM (R(V)=3.1); the reddening of the ionized gas is
thus parametrized by the color excess E(B$-$V)$_{gas}$ (equation~4).

The stellar continuum colors show the trend to become redder for increasing
reddening of the hydrogen emission line ratios (Figure~\ref{fig6}, panels
a--c--d). The implications are twofold: (1) variations of the intrinsic
SED from galaxy to galaxy are secondary relative to the trend induced by dust
obscuration, and (2) within each starburst, the same foreground--like dust
geometry holds for both the ionized gas and the non-ionizing stellar
continuum.

The first implication is not surprising, at least at UV
wavelengths. Since we are observing the Rayleigh-Jeans part of the
massive stars spectrum, the intrinsic UV spectral
slope\footnote{$\beta_{26, o}$ and $\beta_{26}$ refer to the intrinsic
and observed, respectively, UV spectral slope measured between
0.125~$\mu$m and 0.26$\mu$m as defined in \citet{cal94}, see
Appendix~B for more details.} $\beta_{26, o}$ is relatively constant
over a fairly large range of ages, with values
$-$2.7$\gtrsim\beta_{26, o}\gtrsim-$2.2 for constant star formation
over 10$^6$~yr$\lesssim$age$\lesssim$10$^9$~yr \citep[see][and the
Table in Appendix~B]{lei95}. In the same conditions, the production of
ionizing photons is also constant. In an instantaneous burst of star
formation, the ionizing photons are more age-sensitive than the
non-ionizing photons that produce the stellar continuum and disappear
before appreciable changes in the UV spectral shape can be
observed. The observed UV spectral slope $\beta_{26}$ of the
UV-selected starbursts span the range
$-$2.5$\lesssim\beta_{26}\lesssim+$0.4, much larger than what expected
from stellar population variations alone, and increases for increasing
values of the attenuated-to-intrinsic hydrogen line ratios (i.e.,
color excess, Figure~\ref{fig6}).

In addition to correlate with the reddening of the ionized gas, the
reddening of the UV stellar continuum, as measured by $\beta_{26}$,
correlates with the the infrared-to-blue and the infrared-to-UV
luminosity ratios, L$_{IR}$/L$_B$ and L$_{IR}$/L$_{0.16}$
\citep[(Figure~\ref{fig7} and][]{cal95b,meu99}. These ratios are a
measure of the total dust opacity in the starburst region, because the
bulk of the starburst's energy is coming out in the UV--B. In
addition, dust {\em absorption} is the main process that removes light
from the line of sight \citep{cal95b}; when large galactic regions are
observed light scattered by dust out of the line of sight compensates,
on average, that scattered into the line of sight. The
opacity--reddening correlation means that measurements of reddening
can be used to infer the total dust absorption in UV-selected
starbursts \citep{meu99} with an uncertainty,
$\Delta$A$_{0.16}\sim$0.6~mag for individual objects and $\sim$20\%
when averaged over large samples \citep{cal00}, that is dominated by
variations in the starburst populations and in the details of the dust
geometry from galaxy to galaxy. Such a correlation is expected for a
foreground-like dust geometry \citep[Figure~\ref{fig3}, panel~d; see
discussion in][]{gor00}.

The slopes of the correlations in Figure~\ref{fig6} identify the reddening
suffered by the stellar continuum, A$_{\lambda1, star}-$A$_{\lambda2,
star}$=[k$^e$($\lambda_1$)-k$^e$($\lambda_2$)]E(B$-$V)$_{gas}$, with
k$^e$($\lambda$) the obscuration curve, while the total obscuration
measurements of Figure~\ref{fig7} provide the zeropoint for k$^e$($\lambda$)
\citep[see, also,][]{cal00}. The intrinsic starburst flux density
f$_i(\lambda)$ is thus recovered from the observed flux density f$_o(\lambda)$
via:
\begin{equation}
f_i(\lambda) = f_o(\lambda)\ 10^{0.4 E(B-V)_{gas}\ k^e(\lambda)},
\end{equation}
where the obscuration curve for the stellar continuum, k$^e$($\lambda$), is
given by \citep{cal00}:
\begin{eqnarray}
k^e(\lambda) &=& 1.17\, (-1.857 + 1.040/\lambda) + 1.78 \ \ \ \ \ \ \ \ \ \ \ \ 0.63\ \mu m \le \lambda \le 2.20\ \mu m \nonumber \\
           &=& 1.17\, (-2.156 + 1.509/\lambda - 0.198/\lambda^2 + 0.011/\lambda^3) + 1.78 \nonumber \\
           & &\ \ \ \ \ \ \ \ \ \ \ \ \ \ \ \ \ \ \ \ \ \ \ \ \ \ \ \ \ \ \ \ \ \ \ \ \ \ \ \ \ \ \ \ \ \ \ \ \ \ \ \ 0.12\ \mu m \le \lambda < 0.63\ \mu m.
\end{eqnarray}
The above expression folds into a single functional form a variety of
effects: extinction proper, scattering, and the geometrical
distribution of the dust relative to the emitters. It is derived from
the spatially integrated colors of the entire stellar population in
the starburst, and represents the `net' obscuration of the population
itself. As mentioned above, the obscuration curve has mainly an
absorption component, as the effects of scattering are averaged
out. Thus, it should not be confused with the extinction curve
k($\lambda$) defined in section~2.1. Equation~8 implies that 1~mag of
reddening between B and V for the ionized gas, A$_{B, gas}-$A$_{V,
gas}$=1~mag, corresponds to A$_{B, star}-$A$_{V, star}$=0.44~mag of
reddening for the stellar continuum. This fact can be alternatively
expressed as:
\begin{equation}
E(B-V)_{star} = 0.44 E(B-V)_{gas}.
\end{equation}
In other words, the stellar continuum suffers roughly half of the
reddening suffered by the ionized gas \citep[][see, also,
Figure~\ref{fig3}, panels~e and f]{cal94,fan88,mash89}. Although
E(B$-$V)$_{star}$ parametrizes the amount of reddening of the stellar
continuum, it is not a color excess proper as in the case of
individual stars, because of the dust geometry effects folded in the
expression of k$^e$($\lambda$). One obvious characteristic of the
obscuration curve is the absence of the 0.2175~$\mu$m bump, which is a
prominent feature of the MW extinction curve
(Figure~\ref{fig1}). Dust/emitter geometries that can dilute the bump
include mixed distributions with large optical depths
\citep{natta84,gra00,gor00}. However, mixed geometries do not account
for the full range of observational data (Figure~\ref{fig3}), because
the sources that contribute to the emerging SED are located at
progressively lower optical depths $\tau_V$ for decreasing
wavelength. In particular, the reddest UV slope produced by mixed
geometries is $\beta_{26}\sim -0.7$ \citep{charlot00}, to be compared
with $\beta_{26}\sim 0$ observed in starburst galaxies
(Figure~\ref{fig6}, panel~a). Thus, the lack of the feature is
probably intrinsic to the {\em extinction curve} of the starbursts
\citep{gord97}, and may be due to the high UV energy densities that
characterize these regions (sections~2.1 and 4.2).

\subsection{The Physical Origin of the Dust Geometry in `UV--bright' Starbursts}

This section describes a dust model that attempts to reconcile two
apparently contradictory facts: (1) stars in starbursts are on average
a factor $\approx$2 less reddened than the ionized gas (equation~9),
and (2) both stars and ionized gas are affected by the same
foreground-like dust distribution (Figure~\ref{fig6}). A schematic
representation of a dust/star/gas configuration that can at the same
time account for both effects is shown in Figure~\ref{fig8}. Stars are
born within optically thick molecular clouds, and the short-lived,
most massive stars remain closely associated with their parental cloud
and the gas they ionize for their entire lifetime
\citep[e.g.,][]{walb99}. These stars and their surrounding gas will be
generally highly attenuated. As the starburst population evolves, the
inside of the region of star formation becomes depleted of dust
\citep{cal96}. With energy densities $\gg$100 times higher than in the
local ISM, starburst environments are rather inhospitable to
dust. Shocks from supernovae destroy dust grains, via grain-grain
collisions and sputtering \citep{dra79,jon94}. The average lifetime of
a refractory grain is $\sim$8$\times$10$^6$~yr, for a supernova rate
of 0.05~yr$^{-1}$ and an ISM mass of 5$\times$10$^8$~M$_{\odot}$
\citep[e.g.,][]{mck89}, about one order of magnitude or more shorter
than the typical lifetimes of starbursts
\citep{cal97,cal97b,greg98,aloi99}. In addition, hot-star-winds- and
supernova-driven outflows evacuate both interstellar gas and dust from
the region \citep{heck90,deyou94,macl99,ferra00,heck00}. R136, the
central cluster in 30~Doradus, is an example of a few million years
old cluster surrounded by an evacuating region
\citep{scowe98}. Ionized gas separated, in projection, from the hot
stars has also been observed in nearby starbursts
\citep{hunter97,wang98,cal99b,mai00}.

The evacuated (clumpy) dust will act as a foreground-like distribution
for both the gas (also at the edges of the region) and for the central
stars \citep[Figure~\ref{fig8}, ][]{cal96,wit00,gor00}; however, the
ionized gas will be more obscured than the stars, because of its
spatial location and closer association with the dust
\citep{cal97}. This gas is more likely to contribute to the nebular
emission observed at UV--optical wavelengths than the gas in molecular
clouds, because of its milder obscuration. In the presence of multiple
generations of starburst populations, the long-lived, non-ionizing
stars have time to `diffuse' into regions of lower dust density
\citep{cal94,charlot00,gra00}, as their native clusters are disrupted
by evaporation or by the host galaxy's gravitational field
\citep{leisa88,kim99,tre01}. Cool stellar populations in the LMC have,
for instance, been observed to be less embedded in dust than hotter
stellar populations \citep{zari99}. A similar trend is observed in the
young ($<$20~Myr) open clusters of the Milky Way
\citep{yada01}. Massive, non-ionizing stars can still produce
significant UV emission. Hence, the integrated UV--optical stellar
continuum from the combination of the ageing, diffusing populations
and the starburst-embedded young populations will be less obscured
than the emission from the nebular gas \citep{cal94,gasd96}. For the
stellar continuum and the ionized gas to have correlated obscuration
values (equation~9), multiple starburst generations or long-lasting
star formation events (rather than `instantaneous' events) are needed
to attain an equilibrium between the number of stars drifting out of
molecular clouds/starburst regions and the number of newly-born, dusty
stars. Starbursts hosted in dwarf galaxies are observed to have
on-going star formation over $\gtrsim$200~Myr
\citep{cal97b,greg98,aloi99}, and durations could be as long in more
massive starburst galaxies \citep{cal97}.

Although the obscuration curve can be widely applied to UV-selected
starbursts, it is not necessarily applicable to systems with different
stellar population or ISM characteristics. In the case of
non-starburst galaxies there is no obvious mechanism for creating a
foreground-dust-like geometry.  In extreme starbursts as the ULIRGs,
the non-applicability of the obscuration curve has been demonstrated
observationally \citep[][and Figure~\ref{fig7}, panel~a]{gold01}. In
these galaxies, the high central concentration of gas and dust is
likely to induce more extreme geometries than in UV-selected galaxies,
and to produce combinations of clumpy mixed and foreground
distributions. Galaxies whose infrared properties are intermediate
between the ULIRGs and the UV-selected galaxies will progressively
deviate from the dust geometry of the latter
\citep{alon01,forst01}. Dust in subsonic HII~regions is likely to be
mixed with the stars, rather than having been pushed away or destroyed
as in starbursts, and scattering of stellar light into the line of
sight may be an important component in the attenuation budget
\citep{capla86,bgkz01}. Because of the differences in the physical
environment between the two types of objects, the foreground-dust-like
approximation no longer holds for subsonic HII regions and equation~8
does not apply. Differences in the star formation history can also
contribute to the difference between subsonic HII regions and
starbursts, with instantaneous bursts being a better description for
the former and extended star formation for the latter; thus,
differential attenuation between ionized gas and stars is not
necessarily expected in the HII regions. Supersonic HII regions could
be at the boundary between the two extremes just discussed; one
example is 30~Doradus, that has evolved to the point of resembling a
mini-starburst \citep{walb91}.

To conclude, the starburst obscuration curve is a purely empirical
result, and its derivation is independent of any assumption about the
starburst's population, the dust geometry, or the details of the
extinction curve. It provides statistical estimates of the reddening
and obscuration of local UV-selected starbursts, and deviations as
large as $\sim$0.6~mag from the mean obscuration at 0.16~$\mu$m are to
be expected on an case-by-case basis. The physics of the starburst
environment provides an explanation of the `foreground-dust-like'
behavior of the curve, of the correlation between the reddening of the
ionized gas and that of the stellar continuum, and of the correlation
between reddening and total obscuration.  As a sample, local
UV-selected starbursts are characterized by modest amounts of
reddening and obscuration, that are correlated with the star formation
activity (equation~6). The median color excess is
E(B$-$V)$_{gas}\sim$0.35 \citep{cal97c}, corresponding to UV continuum
attenuations of A$_{UV, star}\sim$1.6,1.4,1.1~mag at
0.15,0.20,0.30~$\mu$m, respectively \citep[Table~\ref{tbl-4}
and][]{bua98,meu99}.

\subsection{Useful Expressions}

Observables like the UV slope, the colors, and the luminosities and
equivalent widths of the hydrogen lines can be readily related
to the dust-induced color excess of the ionized gas and, through
equation~9, of the stellar continuum. The UV slope is correlated with 
E(B$-$V)$_{gas}$ (Figure~\ref{fig6} panel~a), and equation~8 gives:
\begin{equation}
\beta_{26}=2.08~E(B-V)_{gas} + \beta_{26, 0}, 
\end{equation}
where a list of $\beta_{26, 0}$ values is given in the Table of
Appendix~B. The slope of the correlation between $\beta_{26}$ and
E(B-V)$_{gas}$ is somewhat steeper than that measured by \citet{cal94}
directly from the data (1.88$\pm$0.28), but the two are within
1~$\sigma$ of each other. It should also be noted that equation~10
becomes a lower envelope to the data points in Figure~\ref{fig6},
panel~a, for $\beta_{26, 0}<-$2.25, a value appropriate for
dust-free, star-forming populations. This indicates that variations of
the dust/stars/gas geometry from the schematic model of
Figure~\ref{fig8} and/or of the stellar population from the
theoretical extreme $\beta_{26, 0}\le-$2.25 has the effect of
reddening the UV slope even for E(B-V)$_{gas}\sim$0. The vertical
scatter in Figure~\ref{fig6}, panel~a should thus be interpreted as
due to such variations from galaxy to galaxy.

Reddening expressions for optical colors can be directly derived from
equation~8 (see, also, Figure~\ref{fig6}, panels~c and d). A
mixed UV--optical color, 0.28$-$V, could be useful to infer the
reddening of intermediate redshift ($<$z$>\sim$0.6) starburst galaxies
observed in B and I, and the reddening of one relative to the other is 
given by:
\begin{equation}
A_{0.28, star}= 1.80 A_{V, star}.
\end{equation}
Under the assumption that the restframe UV emission and the nebular
lines are both due to massive stars, the reddening of the UV continuum
relative to the nebular hydrogen emission is given by:
\begin{eqnarray}
A_{0.16, star}= 1.21 A_{H\beta, gas}\\
A_{0.16, star}= 1.78 A_{H\alpha, gas}\\
A_{0.28, star}= 1.29 A_{H\alpha, gas}.
\end{eqnarray}
Figure~\ref{fig6}, panel~e shows the comparison between data of local
galaxies and equation~(13) \citep{cal97c}. While the agreement between the
data in the first four panels of Figure~\ref{fig6} and equation~8 is a
direct consequence of the fact that those data were used to derive the
starburst obscuration curve, the match between data and expected trend in the
bottom two panels of the figure is a consistency check, since it was not
imposed a priori.

One immediate consequence of the differential reddening between gas and stars
is that the equivalent widths of the nebular lines depend on the amount of gas
obscuration; for the H$\alpha$ and H$\beta$ emission lines the relations are
\citep{cal94}:
\begin{eqnarray}
\log [EW(H\beta)_{a}/EW(H\beta)_{o}]= -0.64 E(B-V)_{gas},\\
\log [EW(H\alpha)_{a}/EW(H\alpha)_{o}]= -0.40 E(B-V)_{gas},
\end{eqnarray}
where the attenuated and intrinsic EWs are directly compared. Comparison
of these trends with the data points is shown in panel~f of
Figure~\ref{fig6}; the trend given by the obscuration curve marks an
upper envelope to the ensemble of observed EWs, because the presence
of older stellar populations underlying the starbursts dilutes the
contrast between the emission lines and the stellar continuum in the data.

Statistically speaking, the UV slope is a good estimator of the total
obscuration affecting the starburst \citep[][and
Figure~\ref{fig7}]{meu99}; in particular, the ratio of the dust
infrared luminosity, L$_{IR}$, to the stellar UV or blue luminosity,
L$_{0.16}$ or L$_B$, is related to $\beta_{26}$ via:
\begin{eqnarray}
\log \biggl[{1\over 1.68} \biggl({L_{IR}\over L_{0.16}}\biggr)_{star} + 1\biggr] \simeq
{A_{0.16, star}\over 2.5} = 0.84 (\beta_{26}-\beta_{26, 0}),\\
\log \biggl[{1\over 3.85} \biggl({L_{IR}\over L_B}\biggr)_{star} + 1\biggr]\simeq {A_{B, star}\over 2.5} =
 0.43 (\beta_{26}-\beta_{26, 0}).
\end{eqnarray}
The two constant values on the left-hand-side of equations~17 and
18, 1.68 and 3.85, represent the bolometric corrections of the
stellar emission relative to the UV or B. The correction for the
B-band is larger and subject to more uncertainties than that for the
UV, hence the larger scatter in the data point of Figure~\ref{fig7},
panel~b. Equation~17 can be rewritten as:
\begin{equation}
\log\biggl[{1\over 1.68} \biggl({L_{IR}\over L_{0.16}}\biggr)_{star} + 1\biggr] = 1.76 E(B-V)_{gas},
\end{equation}
where the equation above represents a lower envelope to the data
\citep[Figure~(7), panel~d and][]{cal00}, and has a scatter
comparable to that of Figure~\ref{fig6}, panel~a.  The nature of the
scatter in the $\beta_{26}$ vs. E(B-V)$_{gas}$ and in the
L$_{IR}$/L$_{0.16}$ vs. E(B-V)$_{gas}$ relations is the same: it is
due to variations from galaxy to galaxy of the starburst populations
and of the dust/star/gas geometry.

\section{Measurements of Dust Opacity in High-Redshift Galaxies}

As discussed in the Introduction, the energy balance between UV--optical and
infrared cosmic background suggests that there have been infrared luminous
galaxies at all epochs. The number density of quasars was $>$100 times higher
at z$\sim$2.5 than today \citep[e.g.,][]{sha99} and star formation activity
was $\sim$3 to 10 times higher at z$\gtrsim$1 than at z$\sim$0
\citep{lil96,cow99}, flattening or decreasing only beyond z$\sim$1--2
\citep{mad96,ste99}. If the association between activity and dust opacity
observed in the local Universe (equation~6) holds at high redshift as well,
galaxies in the past may have been as dusty as or dustier than today
(section~5.2).

Among the methods described in section~3, the multiwavelength comparison is
the only practical approach for measuring the dust opacity of galaxies beyond
the local few hundred Mpc. However, because of observational difficulties, the
wavelength coverage of the typical sample is limited to a few windows, often
with little overlap from sample to sample. Measures of the content, nature,
and opacity effects of the dust in distant galaxies are therefore still at a
very early stage, and the picture is vastly incomplete and controversial.

\subsection{Dust in Damped Ly$\alpha$ Systems}

The very first evidence for presence of dust in high redshift systems
came from studies of Damped Ly$\alpha$ Systems
\citep[DLAs,][]{mey87,fal89,mey90,pet90,pei91,pet94,zuo97,pet97,vla98,pet00,ge01}. DLAs
are the largest HI column density absorption systems seen in the
spectra of background quasars, with
N(HI)$\ge$10$^{20.3}$~cm$^{-2}$. Their metallicities cover the range
$\sim$1/100--1/10~solar for z$\sim$0.5--4, with a mean value
$\sim$1/13~solar at z$\sim$0.5--3 and little evidence for evolution with
redshift \citep{pro00,pet99b,pet972,lu96}. Selection biases against
the dustiest, and most metal-rich DLAs and problems with dust
depletion corrections have been advocated as possible reasons for the
observed weak metallicity evolution \citep[][see, however,
\citet{ellis01}]{fal93,boi98,sava00}. The dust fraction in DLAs is
measured by comparing the abundance of dust-depleted metals such as
Cr~II, Fe~II, and Si~II against that of a metallicity tracer such as
Zn~II, all of them producing rest-frame UV absorption lines
\citep{mey87}. The dust-to-metals ratio in DLAs is in the range
50\%--60\% of the Milky-Way value; this, combined with the generally
low metallicities, indicates that DLAs are relatively transparent
systems, with dust-to-gas ratios between 2\% and 25\% of the Milky-Way
value \citep{fal89,pei91,pet97,vla98}. The absence of the
0.2175~$\mu$m dust absorption associated with DLAs led \citet{pei91}
to conclude that the extinction curve in these systems is probably
similar to the one observed in the Magellanic Clouds. More recently,
though, \citet{mal97} reported a $\sim$3~$\sigma$ detection of the
dust absorption feature in the composite spectra of MgII
absorbers. Although presence of modest quantities of dust is well
established in DLAs, the weak or absent evolution of the metal content
and of the cosmological mass density with redshift \citep{rao00}
questions their relationship to luminous galaxies, and favors the
interpretation that DLAs preferentially trace galaxies with extended
gas distributions and low metal and dust contents \citep{leb97,turn01}
and the unevolved outskirts of galaxies \citep{fergus98}.

\subsection{Dust Emission from Distant Star-Forming Galaxies}

In recent years, facilities/instruments like ISO and the
Sub-millimeter Common User Bolometer Array (SCUBA) at the James Clerk
Maxwell Telescope (JCMT) have achieved high enough sensitivities in
the mid-- and far--infrared to detect {\em dust emission} beyond a few
$\mu$m from statistically significant samples of galaxies at
cosmological distances. Studies on the ISO and SCUBA surveys,
especially the identification of the optical/near--IR counterparts to
the infrared sources and the redshift measurements, are still ongoing and
the results described below should be considered in many instances
preliminary.

ISO surveys have yielded $\approx$1,500 sources in the redshift range
0$\lesssim$z$\lesssim$1.5, down to a sensitivity limit of $\approx$0.1~mJy in
the mid--IR (6.75~$\mu$m, 12~$\mu$m, and 15~$\mu$m) and $\approx$0.1~Jy in the
far--IR \citep[90~$\mu$m and
175~$\mu$m,][]{row97,oli97,tan97,kaw98,cle99,alt99,aus99,flo99a,flo99b,pug99,ser00,sco00,efs00,dol01}.
Full lists of all surveys are given in \citet{rowan99} and \citet{elb99a}. The
number counts at 175~$\mu$m account for $\sim$10\% of the CIB
\citep{pug99,dol01}, showing that the bulk of the background has not been
resolved yet at these wavelengths. Neverthless, the number counts at mid-- and
far--IR wavelengths are about 3--10 times above the no-evolution extrapolation
of the local IRAS counts \citep{elb99b,aus99,pug99,ser00,efs00}. The
implication is a strong luminosity evolution, L$_{IR}\propto$(1$+$z)$^{4.5}$,
or a combination of luminosity and density evolution of the infrared sources
out to z$\sim$1 \citep{cle99,xu00,char01,fra94,gui98}, that is comparable to
or stronger than the luminosity evolution observed at restframe UV wavelengths
out to the same redshift \citep{lil96,cow99,sul00}. The ISO counts are thus
dominated by luminous infrared galaxies at z$\lesssim$1, with
L$_{bol}\sim$L$_{IR}\gtrsim$10$^{11}$~h$_{65}^{-2}$~L$_{\odot}$
\citep{row97,elb99a,xu00}. Although this conclusion is based on the highly
uncertain conversion of the 15~$\mu$m emission\footnote{At z$\sim$1 the flux
density at 15~$\mu$m is centered on the restframe emission of the aromatic
features that dominate the region between 6~$\mu$m and 12~$\mu$m; the relation
between the intensity of the features and the total infrared emission is still
unsettled \citep{bos98,dal01}.} to a total infrared luminosity, it is
qualitatively supported by the optical counterparts of the ISO sources. Close
to 100 mid--IR sources, with median redshift $<$z$>\sim$0.7--0.8 have been so
far identified at optical/near--IR wavelengths \citep{aus99,flo99b,rig00}
[$<$z$>\sim$0.1--0.2 for the 12$\mu$m sample \citep{cle01}]. Many of the
restframe optical spectra show characteristics of dusty luminous starbursts,
with current bursts of star formation superimposed on a prominent population
of A stars, and with attenuation values A$_V\gtrsim$0.5--1.8~mag
\citep{flo99b,rig00,pog00}.

At the time of writing, SCUBA surveys at 850~$\mu$m of `blank' fields and
targeted lensed objects have detected a total of $\approx$100 sources with
flux densities in the range $\sim$0.5~mJy to $\gtrsim$10~mJy and redshifts
between z$\sim$1 and 3, although sources as far away as z$\approx$5 may be
present
\citep{sma97,hug98,bar98,sma98,eal99,lil99,sma99,bar99b,bar00,sma00,eal00,chap01,fox01,scott01}.
The number counts down to 2~mJy and to 0.5~mJy account for $\sim$20\%--30\%
and more than 90\%, respectively, of the CIB
\citep{hug98,bar98,eal99,bar99a,eal00,bla99b}. However, the latter fraction is
potentially affected by small number statistics: so far only a few sources
have flux densities below 2~mJy, the SCUBA confusion limit in blank fields,
all of them from the survey of lensed galaxies \citep{bla99a}. Observations at
850~$\mu$m probe the Raileigh-Jeans tail of the thermal dust emission of
galaxies at least to redshift $\approx$5--7. A source at z$\ge$1 and with an
850~$\mu$m flux density of 2~mJy corresponds to a luminosity
L$_{IR}\sim$2$\times$10$^{12}$~h$_{65}^{-2}$~L$_{\odot}$, for typical SEDs of
local infrared--selected galaxies \citep{lis00,dun00,dun01}; this is roughly
the luminosity of a ULIRG like Arp220. The few optical/near--IR counterparts
which have been identified are faint, often with I$>$24 and K$\gtrsim$20--21
\citep{sma98,bar99b,sma99,lil99,ivi00}; in most cases the ratio
L$_{IR}$/L$_R<$0.05, where L$_R$ is the luminosity in the observer--frame
R~band, and the infrared emission accounts for more than 95\%--99\% of the
bolometric luminosity. It is indeed widely accepted, also on the basis of the
few available multi-wavelength SEDs, that the SCUBA sources are `cosmological
ULIRGs' \citep{hug98,bar98,lil99,sma98,ivi00}, although this generalization
has been recently questioned \citep{eal00}. The median redshift of the surveys
is $<$z$>\simeq$2.5 \citep{lil99,sma99,bar99b,bar00,eal00,sma00,chap01,fox01};
however, $<$10\%~ of all the sources have reliable redshift determinations
from optical/near--IR counterparts, and a total of $\approx$50\%~ of the
sample have photometric redshifts, most often derived from the radio/sub-mm
spectral-index relation of \citet{car99}. The comoving number density of SCUBA
sources in the redshift range z=1--3 brighter than 6~mJy is
$\sim$3--5$\times$10$^{-5}$~h$_{65}^3$~Mpc$^{-3}$ in our cosmology
\citep{bar00,scott01}, about two orders of magnitude larger than the
corresponding number density of local ULIRGs brighter than
10$^{12}$~L$_{\odot}$ \citep{kim98}.  A strong luminosity evolution, possibly
as strong as L$_{IR}\propto$(1$+$z)$^{3}$ to z$\sim$2, is also implied by the
number counts \citep[e.g.,][]{scott01}. Although the statistical properties of
the sources are often inferred using the star-forming Arp220 as a template, a
fraction of AGNs is present in the sample \citep{ivi98,bar99b}. Current best
estimates, based on the X-ray luminosity of the sources favor an AGN fraction
$\sim$20\% \citep{sev00,horn00,fab00,bar01}; this value will generally depend
on the source luminosity range, as it does in the local Universe
\citep{gen98,lut98,vei99}.

\subsection{Dust Reddening and Obscuration in Distant Star-Forming Galaxies}

Multi-wavelength observations of optically-selected samples of
intermediate and high redshift galaxies are generally limited to the
rest-frame UV and optical windows, with few exceptions
\citep[e.g.,][]{flo99b,cha00}. Such surveys suffer the same limitations
of local UV--optical samples, with the added complication of the often
incomplete wavelength coverage. By probing only the {\em differential
attenuation}, the UV--optical data provide, in principle, lower
limits to the actual dust opacity of the galaxy (section~2.2, but see
also section~4). In order to differentiate an ageing stellar population
from dust as the cause of the SED reddening, a variety of approaches
have been adopted by different authors. Common diagnostics include the
Balmer line decrement H$\alpha$/H$\beta$, line-to-continuum ratios,
e.g.  H$\alpha$/UV and H$\beta$/UV, and multi-color SED
fitting. Limitations to each of these diagnostics are discussed in
section~2.2 and in the sections below, as appropriate.

\subsubsection{Galaxies at redshift z$\lesssim$2}

For a UV-selected sample of star-forming galaxies at z$\approx$0.2,
\citet{sul00} and \citet{sul01} derive an average attenuation
A$_{H\alpha}\sim$0.8~mag and find tentative evidence that those
systems follow an obscuration--SFR correlation similar to that
observed in more local galaxies \citep{wan96,heck98,hopk01}. The
authors use a combination of UV (0.20~$\mu$m), optical (H$\alpha$ and
H$\beta$) and radio (1.4~GHz) data to analyze their sample, although
their reddening estimates are mostly based on the subsample for which
they can measure H$\alpha$/H$\beta$ ratios. If the starburst
obscuration curve (section~4) applies to z$\approx$0.2 star-forming
galaxies, an attenuation A$_{H\alpha}\sim$0.8~mag for the nebular line
corresponds to A$_{0.20}\sim$1.3 and A$_{0.16}\sim$1.4 for the UV
stellar continuum. Although these figures agree with local
measurements of the extinction at UV and H$\alpha$ in star-forming
galaxies (section~3.6 and Table~\ref{tbl-3}), the UV selection of the
sample potentially biases the reddening measures towards low values by
excluding more heavily extincted galaxies.

The z$\sim$1 Balmer--Break sample of \citet{ade00} is also a
restframe-UV/blue--selected one. Exploiting the presence of the Balmer
discontinuity at $\sim$0.365~$\mu$m in galaxy SEDs to select z$\sim$1
candidates (a derivative of the Lyman-Break technique, see next section),
these authors have obtained spectroscopic confirmation for $\approx$700
galaxies. The sample contains a large variety of galaxies, from active
starbursts to quiescent star-forming to post-starbursts. If the z$\sim$1
galaxies with blue ($\beta_{26}\lesssim-$0.3) SEDs are analogs of local
starbursts and follow the same IR/UV-versus-$\beta$ relation, they also follow
an obscuration--SFR relation similar to the local one, with a comparable range
of attenuations (Figure~\ref{fig9}, panel~a). However, for fixed opacity,
the z$\sim$1 galaxies tend to have on average brighter UV$+$IR luminosities
(larger SFRs) than the z$\sim$0 galaxies. Thus, even taking selection effects
into account, the distant galaxies are, on average, more actively star-forming
for the same amount of obscuration than the local starbursts \citep{ade00}.

\citet{gla99}, \citet{yan99}, and \citet{moor00} use restframe
H$\alpha$ observations and H$\alpha$/UV comparisons to infer dust
reddening in star-forming galaxies at redshift z$\sim$1--2.2. The
I-band--selected sample of \citet{gla99} corresponds to restframe
$\sim$V for a z=1 galaxy: the longer wavelength selection of the
samples mitigates, although does not remove completely, the bias
against dust obscured galaxies introduced by UV selections [for the
Canada-France Redshift Survey employed by \citet{gla99}, see
\citet{lil95}, and for the sample used by \citet{yan99}, see
\citet{mcc99}]. The presence of dust reddening in the z$\sim$1--2.2
samples is derived from the discrepancy between the measured
L$_{H\alpha}$/L$_{0.28}$ luminosity ratio and the ratio predicted by
dust-free models\footnote{\citet{moor00} measure the restframe
L$_{H\alpha}$/L$_{0.25}$ ratio; the small extrapolation to
L$_{H\alpha}$/L$_{0.28}$ is done here assuming a constant star
formation population.}. Figure~\ref{fig10} compares the range of
attenuated--to--intrinsic values
(L$_{H\alpha}$/L$_{0.28}$)$_{a}$/(L$_{H\alpha}$/L$_{0.28}$)$_{o}$
derived from data by the three groups of authors\footnote{The intrinsic ratio
(L$_{H\alpha}$/L$_{0.28}$)$_{o}$ is an average of H$\alpha$ and UV
luminosities for a variety of IMFs, star formation histories, and
metallicities of stellar populations, see Table~3 in \citet{gla99}.}
with predictions from the dust models of section~2.2. Geometries where
dust and stars are homogeneously mixed, either in spheroidal or
flattened distributions, appear less effective in reproducing the
observed H$\alpha$/UV ratios than foreground geometries
(Figure~\ref{fig10}). If the foreground and starburst distributions
bracket the possible range of dust/emitter morphologies for the
z$\sim$1--2 H$\alpha$ galaxies, the front-to-back optical depths are
$\tau_V\sim$0.6--4.9~mag (Figure~\ref{fig10}, bottom panel), and the
effective attenuations are A$_{H\alpha}\sim$0.6--4.3~mag,
A$_{0.28}\sim$1.5--5~mag, A$_{0.16}\sim$3--7~mag, for SMC
extinction. MW-like and SMC-like extinction curves give results that
are less than 0.1~mag apart at $\lambda\ge$0.28~$\mu$m, but
$\sim$1.2~mag different at $\lambda$=0.16~$\mu$m, with larger
obscurations given by the SMC extinction curve. Uncertainties in the
metallicity, IMF, and star formation history of the galaxy stellar
populations introduce a factor $\gtrsim$2 additional scatter on the
inferred opacities \citep[Table~3 in][]{gla99}. The major limitation
is, however, that dust opacities measured from the current H$\alpha$
surveys of galaxies are probably not representative of the z$\sim$1--2
population as a whole. By number, \citet{gla99}'s and \citet{moor00}'s
samples are small (13 galaxies and 5 galaxies, respectively);
\citet{yan99}'s sample, albeit 2.5 times larger than \citet{gla99}'s,
does not measure the H$\alpha$ emission on the same galaxies for which
UV is available. For such small samples sensitivity limits induce a
strong selection effect: the observed galaxies will be among the
brightest in H$\alpha$ and, in addition, contamination from
AGN/Seyfert nuclei is not excluded \citep{whitt92}. Therefore, such
surveys sample the high end of the activity and dust obscuration
ranges for optically-detectable systems at z$\sim$1--2 \citep[][and
equation~6]{gla99,sul00}.

In summary, the limited information available on galaxies at
z$\approx$1 suggests that dust opacities cover a range similar to that
of local star-forming and starburst galaxies; they may follow a
similar relation A$_{\lambda}\propto \log$(SFR)$^{\alpha}$
(see equation~6), with the proportionality constant revised downward
relative to the local sample.

\subsubsection{Galaxies at redshift z$>$2}

To date, the surveys that have most efficiently secured large populations of
galaxies beyond z$\sim$2 are those that exploit the Lyman discontinuity at
restframe 0.0912~$\mu$m to identify high-redshift candidates; these are the
so--called Lyman--Break Galaxies
\citep[LBGs,][]{guh90,ste92,ste96a,ste96b,gia96,low97,gia98,ade98,ste99,ade00}. So
far, there are $\sim$1,000 spectroscopically confirmed LBGs at z$\sim$3, and
$\sim$70 at z$\sim$4. The LBG population is a substantial component of the
high redshift ``zoo'': their comoving number density at z=3 is
n(z=3)$\sim$3$\times$10$^{-2}$~h$_{65}^3$~Mpc$^{-3}$ \citep{ste99}, comparable
to the density of local galaxies \citep{mar94}. By selection, i.e. due to the
need of presenting a relatively blue continuum longward of Ly$\alpha$, this
high redshift population tends to be biased against systems with predominantly
old populations, or systems with large amounts of dust extinction. Indeed,
most LBGs show evidence for ongoing or recent star formation activity
\citep{ste96b,pet00b,pet01,pap01}. From a purely phenomenological point of
view, LBGs resemble local UV-selected starburst galaxies (section~4). The
obscuration--corrected SFRs per unit galactic area are
$\sim$0.2--25~M$_{\odot}$~yr$^{-1}$~kpc$^{-2}$, for a 0.1--100~M$_{\odot}$
Salpeter IMF \citep{ste96b,ste99,pap01,ashap01}, corresponding to
$\sim$0.5\%--50\% the maximum SFR per unit galactic area measured in the local
Universe \citep{leh96,meu97}; this is similar to the observed range in nearby
UV-selected starbursts. Restframe UV spectra show a wealth of absorption
features, and often P-Cygni profiles in the CIV(1550~\AA) line \citep{ste96b},
which indicates the predominance of young, massive stars. Nebular emission
lines observed in the brightest LBGs \citep{pet98,tep00,pet01} have
intensities comparable to those of local starbursts \citep{meu99}. The
observed blueshifts in the UV interstellar absorption lines relative to the
systemic redshifts of LBGs have been interpreted as bulk gas outflows with
velocities v$\sim$200--500~km~s$^{-1}$ \citep{pet98,pet00b}, again very
similar to what is observed in nearby starbursts \citep{heck00}. Finally, more
luminous LBGs tend to be dustier \citep[Figure~\ref{fig9}, panel~b and
equation~6;][]{ade00,ashap01}. However, dissimilarities also exist between
the two classes of galaxies. Star formation extends over a large fraction of
the galactic body in LBGs, typically 5--30~h$_{65}^{-2}$~kpc$^2$
\citep{gia96,gia98}, while it is confined within the inner kpc$^2$ in local
starbursts. As a result, LBGs are, on average, more rapidly star-forming than
either z$\sim$0 or z$\sim$1 galaxies
\citep[Figure~\ref{fig9};][]{ade00,ashap01}. The bulk outflows in LBGs have
velocities that are similar to those of local infrared--bright starbursts like
NGC6240 and a few times larger than those of HII galaxies
\citep{heck00,heck97,kun98,gondel98}. The implications of such differences are
not fully understood yet.

Establishing how much dust LBGs contain is the subject of ongoing
debate. Metals are definitely present in these systems. Although the exact
value of the metallicity has not been constrained, it most likely lies
between 0.1~Z$_{\odot}$ and 0.7~Z$_{\odot}$
\citep{pet00b,lei01,pet01}. The UV spectral slopes of LBGs have median
value $\beta_{26}\sim -$1.5 \citep{ade00}, significantly redder than
the value $\beta_{26, 0}\sim -$2.2--$-$2.7 expected for a dust-free
star-forming population (Appendix~B). If interpreted in the same
fashion as the UV slope of nearby UV-selected starbursts, LBGs suffer
from dust obscuration at the level of A$_{0.16}\sim$1--2~mag
\citep{cal97c,pet98,dic98,meu99,ste99,ade00}, and, possibly, more
\citep{meu97,saw98}. However, a number of uncertainties and
systematics limit the usefulness of UV data alone to derive
obscuration estimates. First and foremost, the high redshift galaxy
for which reddening corrections are sought needs to possess active,
on-going star-formation (i.e., not to be a post-starburst); this
appears to be true for the LBGs as a population, but may not be true
on a individual basis \citep{pap01}. Second, for most LBGs the UV
slope is derived from photometry alone; the conversion of G$-\cal{R}$
colors\footnote{The G and $\cal{R}$ filters have central wavelengths
0.483~$\mu$m and 0.693~$\mu$m, respectively \citep{ste92}.} to
$\beta_{26}$ suffers from uncertainties due to the presence of both
intrinsic (stellar lines and Lyman break) and intervening
(Lyman$\alpha$ Forest) absorption in the photometric filters
\citep{ade00,pet01}. Finally, because of the age--dust degeneracy, the
intrinsic UV slope is a free parameter that can lead to $>$1~mag
difference in the derived obscuration value at 0.16~$\mu$m
\citep[compare][]{saw98,ste99}.

Supplementing the UV with restframe optical data provides further leverage for
determining the level of dust reddening in LBGs.  The negligible correlation
between L$_{H\beta}$/L$_{0.16}$ and the UV slope found for a sample of
$\lesssim$15 luminous LBGs\footnote{Measures of dust reddening obtained by
comparing the UV continuum with the [OIII](0.5007~$\mu$m) emission line
\citep{tep00} are not discussed because of the strong dependence of the
nebular line intensity, up to one order of magnitude variation, on the
chemical and physical conditions of the gas \citep{jan00,charlot01}.} supports
the notion that LBGs are subject to modest reddening
\citep{pet98,tep00,pet01}. Although this result is still dominated by the
large scatter induced by both observational and model uncertainties
\citep{pet01}, it is consistent with what is observed in local UV-selected
starbursts\footnote{All considerations are based on the assumption that the
dust in LBGs is distributed like that of local UV-selected starbursts. If,
instead, the most appropriate dust geometry is a homogeneous spheroid or a
face-on flattened disk, the average UV slope $\beta_{26}\sim-$1.5 corresponds
to obscurations A$_{0.16}\sim$1.2~mag and A$_{H\beta}\sim$0.50~mag, and a
differential obscuration between the two of $\sim$0.7~mag; the latter is
probably too large to be reconciled with observational data. Highly inclined
disks are not considered in the obscuration budget, as they would probably be
too dim in the UV to be selected as LBGs.}, for which an obscuration of
1--2~mag at 0.16~$\mu$m corresponds to the modest reddening
A$_{0.16}-$A$_{H\beta}\sim$0.15--0.3 (section~4.3).

Although the age--dust degeneracy cannot be easily broken even with
the addition of long wavelength data, extending the wavelength
coverage of LBGs to the restframe optical has helped delimiting the
locus of allowed combinations in the age--dust plane
\citep{pap01,ashap01}. Using complementary datasets of LBGs in terms
of luminosity range, \citet{pap01} and \citet{ashap01} have
established that there is an inverse correlation between the age of
the stellar populations in a galaxy and the amount of dust reddening
it suffers: the older the galaxy, the lower its reddening. In
agreement with previous results \citep{saw98}, LBGs whose SEDs are
compatible with young ($\lesssim$10$^8$~yr) stellar populations are
also dusty. Whatever dust is present, it is not preferentially
obscuring selected regions of the galaxies, because of the similarity
between the restframe UV and optical morphologies
\citep{bunk99,dic00}. A typical LBG obscuration of
A$_{0.16}\sim$1--2~mag, with a median of $\sim$1.5~mag and a high-end
tail around 4.5--5~mag \citep{ste99,pap01,ashap01}, is currently the
most broadly accepted value.

Despite all the uncertainties in estimating dust obscuration in LBGs, the
inferred opacity values are modest enough that the high redshift galaxies are
expected to be faint sub-mm emitters, with 850~$\mu$m flux densities at or
below the SCUBA detection limit of $\sim$1--2~mJy \citep[][]{ade00,pea00}. Out
of $\gtrsim$10 LBGs targeted with SCUBA, only a couple have been detected
\citep{cha00,van00}. Table~\ref{tbl-5} gives three examples (two LBGs and one
SCUBA source) of highly reddened z$\sim$3 galaxies, plus the average values of
a typical LBG. The three examples are meant to show some of the most
attenuated cases that are also detected at restframe UV. Table~\ref{tbl-5}
demonstrates that the exact prediction of the flux density at restframe
$\approx$200~$\mu$m is heavily dependent on the adopted dust temperature(s)
and emissivity, because long infrared wavelengths only probe the
Raileigh-Jeans tail of the thermal dust emission in galaxies
(Figure~\ref{fig11}). Local galaxies, although extensively used as analogs of
high redshift galaxies, are of limited help in this respect, given the wide
range of properties they show. If the flux density of a local galaxy beyond
60--70~$\mu$m is parametrized by a modified blackbody with dust emissivity
$\propto\nu^{\epsilon}$:
\begin{equation}
f_{IR}\propto \nu^{\epsilon} B(\nu, T),
\end{equation}
ULIRGs and bright IR-selected galaxies tend to be well described by either a
single-temperature component with T$\sim$40~K and $\epsilon\sim$1.5
\citep{lis00,bla99a} or by a warm/cool two-temperature model with $\epsilon$=2
\citep{dun01}; however, there is tentative indication that UV--bright,
metal-poor starbursts have hotter dust SEDs, with T$\sim$50~K and $\epsilon$=2
\citep[Tol1924--416,][]{cal00}. Equation~20, which is used for purely
illustrative purposes, is clearly an over-simplification of the dust SED in
local galaxies, which is characterized by a range of temperatures and, at
restframe $\lambda\lesssim$40--70~$\mu$m, by the emission from non-equilibrium
dust grains \citep[see, also, section~2.1]{sel84,dra85,li01}. In starbursts,
the latter component represents $\sim$25\% of the total infrared
emission. Given the uncertainty of the sub-mm measures and the current poor
understanding of the detailed physics of dust emission in both local and
high-redshift galaxies, the opacity of LBGs is still relatively unconstrained
by sub-mm observations \citep[][and Table~\ref{tbl-5}]{ouc99,saw01}. Further
complications (and uncertainties) arise when the comparison between
predictions and observations of infrared flux densities is made for individual
objects \citep{van00,saw01,baker01}; reddening/obscuration corrections at
UV-optical wavelengths are accurate only in a statistical sense and,
generally, do not apply on a one-to-one basis to individual objects. For
instance, if the dust geometry in a particular system is such that the
scattering out of the line of sight is larger than that into the line of
sight, the infrared energy, which is due to the absorption part of the
extinction only, will be overestimated when derived from UV data.

\subsection{Summary}

There is so far limited overlap between infrared and optical surveys at
intermediate and high redshift
\citep[e.g.,][]{aus99,flo99b,bar99b,rig00,cha00}. Consequently
multi-wavelength coverage from the restframe $\sim$UV--optical to the infrared
is available only for a small number of galaxies. To make matters worse,
observational limitations lead to vastly different selection criteria for the
optical surveys in different redshift bins
\citep{gla99,yan99,moor00,ste96b}. Thus, conclusions about the dust opacity of
galaxy populations as a whole and as a function of redshift are at the moment
based on `reasonable' extrapolations of existing data, rather than hard
evidence.

Available infrared and optical data support the scenario that z$\approx$1
star--forming galaxies cover at least the same range of dust opacities as
local galaxies. The luminosity (or combined luminosity and density) evolution
implied by the ISO counts suggests that luminous infrared galaxies, with
L$_{bol}\sim$L$_{IR}\gtrsim$10$^{11}$~h$_{65}^{-2}$~L$_{\odot}$, were common
at z$\lesssim$1. The range of dust attenuation values at UV wavelengths
inferred for optically-selected z$\sim$1 galaxies is comparable to those of
local disk galaxies and UV--selected starbursts (Figure~\ref{fig9}), although
the distant galaxies tend to be more luminous (more active) on average. Larger
samples with more extended wavelength coverage are needed before a more
complete picture can be drawn in this redshift range. 

The ISM of z$\sim$3 galaxies has been observed to be metal-polluted to at
least $\approx$0.1~Z$_{\odot}$, implying that non-negligible amounts of dust
must be present in the galaxy population at large by z=3 and, in many cases,
earlier \citep{ashap01}. The observed properties of the z$>$1 SCUBA sources
and of the LBGs suggest that both objects are the high redshift analogs of local
actively star-forming galaxies, where dust content and level of activity
determines a continuum of characteristics. The SCUBA sources are at the high
end of such continuum, with powerful activity, both thermal and non-thermal,
and large dust contents that make them similar to local ULIRGs; more than 95\%
of their bolometric light is absorbed by dust. LBGs appear to span a larger
range of the continuum of dust content and activity level, although they tend
to have, on average, lower values than the SCUBA sources. Typical UV
attenuations are A$_{0.16}\sim$1--2~mag, corresponding to 35\%--95\% of the
bolometric light being absorbed by dust.

Assuming the local infrared luminosity function can be applied to high
redshift sources as well, the comoving number density of z$\approx$3 SCUBA
galaxies with 850~$\mu$m flux densities brighter than 2~mJy and 0.5~mJy
(corresponding to $\sim$2$\times$10$^{12}$~L$_{\odot}$ and
$\sim$5$\times$10$^{11}$~L$_{\odot}$ for an Arp220-like SED) would be
$\approx$4$\times$10$^{-4}$~h$_{65}^3$~Mpc$^{-3}$ and
$\approx$10$^{-2}$~h$_{65}^3$~Mpc$^{-3}$, respectively. High-redshift infrared
galaxies as bright as Arp220 are almost 100 times less numerous than LBGs; for
comparison, in the local Universe ULIRGs are more than 4 orders of magnitude
less numerous than optically selected galaxies. At lower 850~$\mu$m
luminosities, z$\sim$3 SCUBA sources are about as numerous as LBGs at the same
redshift, but there could be large overlap between the two samples at the mJy
level. LBGs are, in fact, predicted to occupy the faint end of the SCUBA
850~$\mu$m source counts, with flux densities of a few mJy or less
\citep{ade00}. Thus, although SCUBA sources down to 0.5~mJy at all redshifts
already account for $>$90\% of the CIB, a non-negligible fraction of this, at
least 25\%, could be due to LBGs \citep{pea00,ade00,van00}.

\section{Intracluster and Intergalactic Dust}

Intracluster and intergalactic dust are briefly discussed in
view of their potential importance in the interpretation of the
intermediate and high redshift Universe.  The amount, nature, and
sources of the dust in the intracluster (ICM) and intergalactic (IGM)
media are still a controversial matter, despite the large number of
independent researches tackling the issue.

The metallicity of the present-day ICM is fairly high, about 1/3--1/2 solar,
implying that a non-negligible fraction of the gas in the ICM has been ejected
from the cluster's galaxies \citep{mus96,mus97,voi97}. The high metallicity is
not automatically accompanied, however, by a high dust content of the ICM. The
ICM hot gas effectively destroys dust through sputtering. The typical
timescale for sputtering in a $\gtrsim$10$^7$~K hot ambient gas is
t$_{sput}\approx$10$^6$a/n, where t$_{sput}$ is measured in yrs, a is the size
of the dust grain in $\mu$m, and n is the density of the ICM in cm$^{-3}$
\citep{dra79,dwe90}. In the high-density cores of clusters
(n$\gtrsim$10$^{-3}$~cm$^{-3}$), a 0.1~$\mu$m dust grain survives a few times
10$^8$~yr, with smaller grains being destroyed on shorter timescales than the
large grains \citep{dwe90}. Small grains can survive only outside the cores of
clusters, i.e., beyond the central $\approx$Mpc. Dust needs to be constantly
replenished in the ICM, and the likely sources are enriched winds from the
cluster's galaxies, intracluster stars, and, in dynamically young clusters,
winds from infalling spiral galaxies \citep{dwe90,pop00}. The timescales for
the dust injection in the ICM are of order 10$^9$~yr or longer, implying that
the grain size distribution, and therefore the shape of the extinction curve,
depends mainly on the sputtering efficiency at the selected cluster location
\citep{dwe90}.

Various direct and indirect methods have been employed to detect dust
in the ICM. Deficits in background counts of galaxies, quasars, and
clusters behind galaxy clusters of various richness have been
interpreted as extinction in the range A$^o_B\sim$0.2--0.5, with
patches that can exceed A$^o_B\approx$1~mag
\citep{zwi57,kar69,bog73,boy88,sza89,rom92}. Conversely, reddening
measurements of cluster members \citep{fer93} or background galaxies
\citep{mao95} give upper limits of A$^o_B\sim$0.2~mag. The two methods
thus give contradictory results and neither is completely bias-free: 
deficits in background counts can be due not only to dust, but also to
confusion; low reddening values can be attributed to a gray extinction
curve, greyer than a Milky-Way-like curve, because of the selective
destruction of small grains in the ICM. Tentative infrared
detections of ICM dust emission, using IRAS and ISO data
\citep{wis93,sti98}, are either at low significance level, or have
been argued to be due to emission from cluster members \citep{qui99}.

Presence of dust is well established in the centers of cooling flow clusters
(see discussion in section~3.4), but the issue of its origin is still
controversial. In particular, it is not clear whether this is an extended
component, such as condensates from the cooled cluster gas or a by-product of
star formation triggered by the cooling flow,
\citep{hu83,hu92,fab94,all95,han95,cra99,edg01}, or is directly associated
with the central galaxy, possibly from a merging event that the galaxy
underwent \citep{spa89,spa93,don00}. The Milky-Way-like form of the extinction
curve in the central galaxies has been extensively used as an argument to
support the galactic origin of the dust in cooling flow clusters; as remarked
above, ICM-processed dust would tend to be depleted in small grains and
therefore produce a gray extinction \citep{spa89,spa93,don00}. However, if
soft X-ray absorption is present in the centers of the clusters, dust
absorption in the cooling flow would appear to be the direct physical
explanation \citep{voi95}. The tentative detection of the neutral oxygen edge,
and thus of dust, in the Perseus cluster \citep{arn98} and the widespread
presence of CO emission in low-radio-emitting cooling flow clusters
\citep{edg01} are countered by XMM data that do not support the presence of
large amounts of cool gas in the center of some of the clusters and, 
therefore, do not require a large amount of dust to explain the soft X-ray
spectral shape \citep{tam01,kaa01,pete01}. 

At present times, the presence and nature of dust in the IGM is even more
uncertain than that in the ICM, though it is widely accepted that the IGM is
metal-enriched. Hydrodynamical CDM simulations identify the Ly$\alpha$ forest
as the IGM, i.e., modest baryon density fluctuations in a ionizing background
\citep{cen94,pet95,zha95,her96,the98}. The Ly$\alpha$ forest accounts for a
significant fraction, $>$50\%, of the baryons at z$\approx$3 \citep{rau97};
its metallicity at z$\sim$2.5--3 is $\sim$0.2--1\%~Z$_{\odot}$ down to column
densities N(HI)$\sim$10$^{14.5}$~cm$^{-2}$ \citep{tyt95,son96,ell99}, while
metallicity values at lower column densities remain uncertain
\citep{ell00}. These values could be lower limits to the IGM metallicity, if
there is a `missing metals' problem as pointed out by \citet{ren99} and
\citet{pet99a} on the basis of the metals produced by star formation by
z$\sim$2.5.  The metallicity of the Ly$\alpha$ forest increases for decreasing
redshift, being $\sim$10 times more metal rich at z$\sim$0.5 than at
z$\sim$2.5--3 \citep{barl98}. In the local Universe, the Ly$\alpha$ forest
accounts for only $\approx$20\% of the baryons \citep{pen00} and matter
associated with galaxies and clusters for another $\approx$25\%--30\%
\citep{fuk98,cen99}; thus, $\sim$50\% of the baryons at z=0, and possibly of
the metals, are contained in a yet undetected phase of the IGM, proposed to be
hot gas \citep{cen99}. Various mechanisms have been considered for the metal,
and possibly dust, pollution of the IGM; they include supernova ejecta from
galaxies and/or subgalactic structures, dynamical encounters between galaxies,
ram-pressure stripping, and population~III stars
\citep{ost96,nat97,gne98,fer00,cen01,mad01,agui01}. The different mechanisms
are non-exclusive, although each one could be dominant relative to the other
at specific cosmological epochs.

Early attempts to detect diffuse dust in the IGM investigated the evolution of
the mean quasar spectral index with redshift \citep{wri81,che91}, with the
basic result that presence of IGM dust is negligible for Milky-Way-like
extinction, up to at least z$\approx$2 \citep{che91}. Milky-Way-like IGM dust
should also leave an imprint on the Cosmic Microwave Background, as it would
re-emit in the sub-mm/mm regime the absorbed background radiation
\citep{row79,wri81,bon91}. The tight COBE upper limit to deviations from
blackbody emission \citep{fix96} limits the IGM dust to A$_V\lesssim$0.3~mag
to z=5, with negligible value to z=3 \citep{loe97,fer99}. However, if
sputtering is efficient in the hot IGM, dust has a greyer extinction than
Milky-Way-like \citep[][for a challenge to the sputtering argument see
\citet{sim99}]{agu99}. Gray IGM dust can be present in larger quantities than
Milky-Way-like dust without violating the COBE limits, but its emission would
represent $\gtrsim$75\% of the CIB at 850~$\mu$m \citep{agu00}. Source counts
with SCUBA account already for $>$90\% of the CIB (section~5.2), which
challenges the case for large amounts of gray dust. So far, the strongest
limits have come from observations of cosmological supernovae. For gray dust
with minimum grain size $\sim$0.1~$\mu$m, \citet{agu99}'s model gives
A$_B$-A$_V\lesssim$0.01 and A$_B$-A$_I\sim$0.2 for A$_V\sim$0.2--0.3 to z=0.5;
at z=2, A$_V\sim$0.6--0.7~mag. Using observations of Type 1a supernovae at
z$\sim$0.5 and z$\sim$1.7, \citet{rie00} and \citet{rie01} have placed a limit
of A$_V\sim$0.1--0.15 to the amount of diffuse, gray IGM dust to z=0.5. In
summary, although the actual numbers are still not settled, IGM dust is likely
to produce small attenuations at most redshifts.

\section{Conclusions and Future Developments}

Current observational data indicate that the population of local galaxies is
characterized by modest values of the dust opacity. Typically, 50\% or less of
the total stellar energy is absorbed by dust. The same evidence points out
that the definition of `modest opacity' is wavelength-dependent; for example,
in the UV, at 0.15~$\mu$m, a non--active galaxy can lose up to 80\% of its
energy to dust. Luminous, intermediate--to--late type spirals are dustier than
ellipticals, by $\sim$1.0--1.4~mag in the UV, and than irregulars, by
0.5--0.8~mag. Luminous galaxies are on average more metal--rich and dust--rich
than their faint counterparts. Dust opacity also increases with thermal and
non-thermal activity, with the high-end of the distribution given by the
ULIRGs.

Although the wealth of data and wavelength coverage for local galaxies has
started to uncover the dust opacity of these systems, we are still missing a
few crucial pieces of the puzzle. The largest among these is the one-to-one
relation between absorbed stellar SEDs and dust emission SEDs. At first order,
hotter stars are expected to heat the dust to higher temperatures than cooler
stars \citep{helo86}, but we don't have yet the direct link between individual
dust emission and heating components; we don't know the dependence of the dust
SED on parameters like metallicity, radiation field, gas density, galactic
environment, etc.; and questions still surround the relation between radiation
intensity and PAH emission, despite the progresses made with ISO
\citep{cesa96,gen00,dal01}.  The main reason for these limitations to our
comprehension of the dust absorption--emission relationship is the large
mismatch in angular resolution between UV-to-near--IR observations and current
mid/far--infrared data. While observations probing the direct stellar light
can easily reach sub-arcsecond resolution, existing data on dust emission are
at the arcminute resolution level, especially at wavelengths above
$\sim$40~$\mu$m where the bulk of the energy from dust emerges. Upcoming
infrared missions like SIRTF (the Space InfraRed Telescope Facility), SOFIA
(the Stratospheric Observatory for Infrared Astronomy), and FIRST (now the
Herschel Space Observatory) promise to bridge this gap. They will also push
the wavelength coverage all the way to 600--700~$\mu$m, thus probing the cold
dust emission in local galaxies, which is the repository of most of the dust
mass.

UV-- and optical--selected, intermediate and high redshift, galaxy surveys
carry the inherent limitation that physical parameters directly derived from
the observations are affected by dust in a non-easily quantifiable way. This
is further complicated by the fact that instrumental sensitivity limits favor
detection of the luminous representatives of any galaxy class, that may tend
to have relatively high dust content. Estimates of the SFRs in
intermediate and high redshift galaxies using Balmer lines
\citep{gla99,yan99,moor00} and infrared emission \citep{flo99b,bla99c,bar00}
are generally a few times higher than the analogous estimates using the
restframe UV emission \citep{lil96,mad96,ste99}. As discussed in section~5,
presence of dust in galaxies at cosmological distances is reasonably
established, but the quantification of its effects as a function of redshift
is still an ongoing process. Linked to this is the puzzle of the nature of the
UV--selected and IR--selected sources, and of their relationship: whether they
represent overlapping populations, or the two ends of a continuum of
properties possibly parametrized by activity and dust content is still an open
question.

The absence of full characterization of the dust emission and absorption in
distant galaxies is mostly due to the scant overlap between
optically--selected and infrared--selected surveys. This is due to sensitivity
limitations at both ends of the spectrum, to angular resolution limitations at
long wavelengths, to the small number of objects in some of the samples, and
to the sparse wavelength coverage of the dust emission SED by current
sub-mm/radio instrumentation. SIRTF and FIRST initially, and then ALMA (the
Atacama Large Millimeter Array) and NGST (the Next Generation Space Telescope)
will be able to fill the gaps. In particular the last two instruments, with
their sub-arcsecond resolution and high sensitivities, should mark a
break-through in our understanding of the high redshift Universe and of its
stellar, gas, and dust content up to z$\sim$20.

\acknowledgments

The author is grateful to Mark Dickinson, Megan Donahue, Paul Goudfrooij,
Claus Leitherer, and Mark Voit for insightful comments on specific sections of
this review. She thanks Kurt Adelberger for data on the Lyman Break and Balmer
Break Galaxies, and Max Pettini and Chuck Steidel for the UV spectrum of
MS1512-cB58. She is also indebted with Veronique Buat, Bruce Draine, Tim
Heckman, Nino Panagia, Marcia Rieke, and Max Pettini for a critical reading of
the manuscript, which has greatly improved its content and presentation.



\appendix

\section{Adopted nomenclature for extinction and obscuration}

The dimming effects of light are generally termed `extinction' only in the
simple geometrical case of a point source behind a dust screen (section~2.1),
where the attenuation suffered by the source can be directly related to the
optical depth of the dust screen. In this review, the extinction at wavelength
$\lambda$ is indicated as:
\begin{equation}
A^o_{\lambda}=1.086~\tau_{\lambda},
\end{equation}
(equation~2). The dimming of background galaxies due to the dust in a
foreground galaxy can generally be considered extinction, because the
foreground galaxy can be modeled in first approximation as a homogeneous
screen in front of point sources (section~3.1).

For more general distributions of dust and emitters, the dimming and reddening
of the light is no longer directly related to the extinction or the optical
depth of the dust layer, and is called indiscriminately `effective
extinction', `obscuration', or `attenuation'. The effective extinction
A$_{\lambda}$ is defined as:
\begin{equation}
A_{\lambda}=m_{\lambda, a}-m_{\lambda, o},
\end{equation} 
where m$_{\lambda, o}$ is the magnitude of the dust--free
source and m$_{\lambda, a}$ is the measured (attenuated) magnitude;
A$_{\lambda}$ measures the magnitudes of attenuation suffered by an
extended source at the specified wavelength/band. By definition,
A$_{\lambda}\equiv$A$^o_{\lambda}$ for a foreground, homogeneous,
non-scattering dust screen.

For dust/emitter distributions that do not have spherical symmetry, such as
spiral galaxies, the face-on effective extinction is termed A$_{\lambda, f}$,
while A$_{\lambda}$ indicates the effective extinction averaged over all
inclinations. 

\section{The definition of the UV spectral slope $\beta$}

Throughout this review, two different definitions of the UV spectral
slope, $\beta$ and $\beta_{26}$ have been used. In this context, the
spectral slope is derived from the fit
f($\lambda$)$\propto\lambda^{\beta}$ within a specified wavelength
range, with f($\lambda$) the flux density of the source. The various
definitions reflect historical reasons driven mostly by observational
constraints.

The default $\beta$ adopted in this review (section~2.2) is measured
between 0.125~$\mu$m and 0.30$\mu$m, using a set of wavelength windows
that avoid the strong stellar and interstellar absorption features. An
example of such windows is given in \citet{cal94}, although the ones
used to derive $\beta$ here are better tuned for use with model
spectra rather than with observed spectra, and extend all the way
to 0.3~$\mu$m. This definition of $\beta$ has been adopted to ensure
an appropriate comparison between different extinction curves and dust
models, as in Figures~\ref{fig2} and \ref{fig3}, because the fitting
windows are selected to avoid the strong and wide dust feature at
0.2175~$\mu$m.

The UV slope $\beta_{26}$ used in section~4 is a fit through the 10
windows of \citet{cal94}, and corresponds to deriving the slope in the
wavelength range 0.125--0.26~$\mu$m, using spectral regions mostly
devoid of strong absorption lines. This definition of the UV slope is
observation-driven, as it was used to obtain measurements from IUE
spectra of a local UV-selected starburst sample. The wavelength
range is bluer than that of $\beta$ above, and was chosen by
\citet{cal94} to avoid including in the fits the older stellar
populations underlying the starbursts. These older populations can
contribute to the spectra starting around 0.27--0.28~$\mu$m, when the
fluxes of F-stars become non-negligible.  For reference, a list of
$\beta_{26}$ values for dust-free stellar populations, both for
instantanous burst and for constant star formation, is given in
Table~\ref{tbl-6}.

An convenient definition of the UV-slope for studies of high-redshift
galaxies is $\beta_{18}$, measured roughly between $\sim$0.12~$\mu$m
and $\sim$0.18$\mu$m. At z$\sim$3, the optical window give information
shortward of $\sim$0.2~$\mu$m, thus forcing the UV slope fitting to a
much shorter range than possible with local galaxies. The details of
the actual wavelength range vary with redshift and with the observing
technique (whether spectroscopy or imaging, and depending on the
filters used).

The relation between the three quantities, as given by models of
stellar populations reddened by the starburst obscuration curve of section~4,
is:
\begin{equation}
\beta_{26}=(1.020\pm0.018)\beta-(0.018\pm0.065),
\end{equation}
\begin{equation}
\beta_{18}=(1.156\pm0.013)\beta_{26}+(0.454\pm0.075).
\end{equation}
The error bar following each coefficient refers to the maximum variance
observed by varying a number of parameters in the models and fits: (a) stellar
populations from both \citet{lei99} and \citet{bruz00} models; (b) stellar
populations with constant star formation in the range 5--1000~Myr, and
instantaneous bursts in the age range 5--12~Myr; (c) both 0.2~Z$_{\odot}$ and
Z$_{\odot}$ metallicities; (d) fitting windows that progressively include or
exclude weak absorption lines in the spectra. The latter turns out to be the
main driver of the variance, while changes in the characteristics of the
stellar populations have only a small impact. The relations between the three
slopes would, however, change from equations~(B1)--(B2) if more extreme cases
of stellar population mix, such as the addition of an old (age$>$1~Gyr)
component, were considered; in this case, $\beta$ and $\beta_{26}$ would be
strongly affected, because of the long wavelength baseline.

Of interest for high-redshift studies is the consideration that while
$\beta_{26}$ and $\beta$ follow each other closely, $\beta_{18}$ is
always redder (more positive) than the other two, and the discrepancy
increases for increasing value of the slopes. The reason for this
counter-intuitive behavior is the presence of the well-known `iron
curtain' at long UV wavelengths: this is a wide dip in the
0.23--0.28~$\mu$m range of the UV spectra due to the presence of a
large number of closely spaced Fe absorption lines \citep{lei99}. The
UV slopes fitted with the longer wavelength baseline have windows in
this continuum-depressed region, hence their more negative
values. Increasing dust obscuration affects the short-wavelength slope
more than the long-wavelength ones; hence the trend for $\beta_{18}$
to become progressively redder than $\beta_{26}$ as their values
increase (i.e., as the UV spectra become more dust reddened). 

\citet{meu99} derive an empirical relation between $\beta_{26}$ and a
short-wavelength UV slope $\beta_{SW}$ (close, but not identical, to
$\beta_{18}$) from IUE spectra of the UV-selected starbursts. Their findings,
that $\beta_{26}$=$\beta_{SW}-$0.16, are consistent with the long wavelength
slope being bluer than the short wavelength one. However, they do not measure
an increasingly redder $\beta_{SW}$ for increasing values of both slopes. A 
possible reason is that noise in the IUE data may have prevented the detection
of such effect. In this review, it is assumed that equations~(B1)--(B2) are
valid for UV-selected starbursts.





\clearpage



\figcaption{Examples of extinction curves in local galaxies. The
Milky-Way extinction curve is shown for three different values of
R(V), 3.1 (continuos red line), 5.0 (dashed red line), and 2.0 (dotted
red line) \citep{ccm89,fit99}. The extinction curve of the Large
Magellanic Cloud's 30~Doradus region (dashed black line) and of the
Small Magellanic Cloud's bar (continuous black line) are reported for
R(V)=2.7 \citep{gor98,mis99}. The starburst obscuration curve
(section~4) is shown (blue line) here for a purely illustrative
comparison. The comparison should not be taken at face value, because
the dust obscuration of galaxies is conceptually different from the
dust extinction of stars. The latter measures strictly the optical
depth of the dust between the observer and the star, while the former
expresses a more general attenuation, in that it folds in one
expression effects of extinction, scattering, and the geometrical
distribution of the dust relative to the emitters
(section~4).\label{fig1}}

\figcaption{Predicted reddening and attenuation values for five simple models
of plane-parallel dust/emitter distributions \citep[section~2.2
and][]{natta84,cal94,wan96}: a homogeneous, non-scattering dust screen
foreground to the light source (red squares); a homogeneous, scattering dust
screen foreground to the light source (red circles); a Poissonian distribution
of clumps in front of the light source, with average number $\cal N$=10 (red
triangles); an homogeneous mixture of dust and emitters (black squares); a
homogeneous distribution averaged over all inclination angles, with increasing
star-to-dust scaleheight ratio from UV (and ionized gas emission) to K (black
circles). Both the MW extinction curve for diffuse ISM (continuous lines) and
the SMC-bar extinction curve (dashed lines) are used in the models. The
reddening and attenuation given by the starburst obscuration curve (section~4)
are also shown for comparison (blue line). $\tau_V$ is the front-to-back
optical depth of the dust layer. Panels~a and b show the relation between
$\tau_V$ and the effective extinction A$_V$ and color V$-$I.  Panels~c--f
show the effects of dust reddening on the UV, optical, and near--IR colors of
a stellar population. The default population is a 300~Myr constant star
formation one. U, V, I are standard Johnson magnitudes; K has central
wavelength 2.163~$\mu$m and zeropoint $-$26. m$_{0.15}$ is defined as
$-2.5\log[f(0.15)]-21.1$, where f(0.15) is the flux density at
0.15~$\mu$m. \label{fig2}}

\figcaption{The same models of Figure~\ref{fig2} for additional spectroscopic
and photometric quantities.  The dust optical depth $\tau_V$ is defined in the
previous figure. R$_{\alpha\beta}$ and R$_{\beta\gamma}$ are the
attenuated-to-intrinsic H$\alpha$/H$\beta$ and H$\beta$/Br$\gamma$ ratios,
respectively (equation~3 and Table~\ref{tbl-2}). The UV spectral slope $\beta$
is defined in Appendix~B. L$_{dust}$ is the dust luminosity, corresponding to
the stellar light removed (absorbed and scattered) by dust from the line of
sight. This is generally larger than the dust infrared luminosity, which only
includes the {\em absorbed} fraction of the energy, unless scattering in and
out of the line of sight compensate each other. The latter can be often safely
adopted for large samples of galaxies with random inclinations (exemplified by
the fifth model). L$_{H\alpha}$ is the luminosity of the nebular emission at
H$\alpha$. Continuum luminosities at 0.15~$\mu$m and 0.28~$\mu$m are given as
L$_{\lambda}=$4$\pi$D$^2 \lambda$f($\lambda$), where D is the galaxy distance
and f($\lambda$) is the flux density at wavelength $\lambda$; the reference
wavelengths are chosen to span the UV spectrum while avoiding prominent
absorption lines (including the broad 0.2175~$\mu$m bump in the MW extinction
curve). The large deviation of the starburst reddening curve from the other
models in the last two panels is due to the differential obscuration between
ionized gas and stellar continuum (section~4).\label{fig3}}

\figcaption{The comparison of colors and absolute magnitude between
an ageing instantaneous burst of star formation in the range
1-80~Myr (black line) and a fixed-age (6~Myr) stellar population
subject to increasing dust obscuration (red line). The model used for
the dust obscuration is a foreground clumpy distribution, with optical 
depth range $\tau_V=$0--6.3.\label{fig4}}

\figcaption{The UV to IR flux ratio as a function of the sum of the UV and IR
luminosities for star-forming galaxies (reproduction of Figure~4 of
\citet{wan96}). The UV flux is centered at 0.20~$\mu$m and is defined as
F$_{UV}$=$\lambda$f($\lambda$). The IR emission in this figure (called `FIR')
is the integrated emission in the 40--120~$\mu$m IRAS window, which contains
roughly between 35\% and 60\% of the dust bolometric emission
\citep{dal01}. The ratio UV/IR is a measure of the total opacity of a
star-forming galaxy (see, also, section~4). The plot shows that more luminous
galaxies are more opaque. The symbols indicate: spirals (filled squares),
irregulars (crosses), elliptical/lenticulars (empty squares), amorphous
(stars), unknown (circles). The lines show predictions for a range of
parameters from the model described in \citet{wan96}. Figure reproduced with
permission from the author and the publishing journal. \label{fig5}}

\figcaption{Panels~a, c, and d show UV-to-near--IR colors of
$\sim$50 UV-selected starbursts as a function of the
attenuated-to-intrinsic hydrogen emission line ratio
H$\alpha$/H$\beta$ (filled circles). The relation between
R$_{\alpha\beta}$ and the color excess of the ionized gas
E(B$-$V)$_{gas}$ (section~4.1) is given by equation~4. The starburst
obscuration curve (blue line) is derived from these data. Increasing
values of R$_{\alpha\beta}$, i.e. larger amount of dust intercepted by
the gas, correspond to redder colors of the stellar continuum
\citep[adapted from][]{cal94,cal97}. For the definition of
$\beta_{26}$ see Appendix~B; f(0.16), f(J), and f(K) are flux
densities at the indicated wavelengths. Panel~b shows the
correlation between the attenuated-to-intrinsic ratios of the hydrogen
line pairs H$\alpha$/H$\beta$ and H$\beta$/Br$\gamma$; plotted with
the data points are trend of the dust models of section~2.2 (with the
same coding of Figure~\ref{fig3}). Homogeneous mixtures of gas
and dust, either face-on or inclination-averaged, are excluded by the
data, indicating that there in little internal dust in the starburst
site \citep[adapted from][]{cal96}. Panels~e and f give for both
the luminosity ratio L$_{H\alpha}$/L$_{0.16}$ and the equivalent width
of H$\alpha$ the comparison between data points and expected trend of
the starburst obscuration curve. \label{fig6}}

\figcaption{The dust bolometric luminosity, L$_{IR}$, normalized to the UV and
B stellar luminosities (L$_{0.16}$ in panel~a and L$_B$ in panel~b,
respectively) shown as a function of the UV spectral slope $\beta_{26}$ for
the same starburst sample of Figure~\ref{fig6} (circles) and for nine ULIRGs
\citep[stars in panel~a;][]{tren99,gold01}. For most galaxies, where only
IRAS measurements are available, the infrared luminosity in the 40--120~$\mu$m
window is transformed into a dust bolometric luminosity using the correction:
L$_{IR}$=1.75~L$_{(40-120)}$ \citep{cal00}.  The blue luminosity is that of
the starburst region, from the data of \citet{mcq95} and \citet{stor95}. The
luminosity ratios measure the starburst total dust opacity, while the UV slope
is a measure of the UV reddening. The correlations imply that measuring the
reddening in a UV-bright starburst gives a good handle on its global
obscuration \citep[adapted from][]{cal95b,meu99}. In panel~a the predicted
trend from some of the models of section~2.2 (coded in the same fashion as in
Figure~\ref{fig3}) are reported; in panel~b, only the starburst obscuration
curve is shown. ULIRGs deviate from the behavior of UV-selected starbursts,
and combinations of clumpy distributions and homogeneous mixtures of dust and
stars seem to better account for the data. Panels~c and d show the same
data of panel~b, as a function of the galaxy metallicity and of the
attenuated-to-intrinsic H$\alpha$/H$\beta$ ratio, respectively. Dust opacity
correlates with the metal content in starbursts \citep{heck98}. \label{fig7}}

\figcaption{A schematic representation of the dust/star distribution
in starburst galaxies, that can account for most observational
constraints. The starburst region (center of figure) contains the
newly formed stellar population, including the short-lived, massive
stars (dark starred symbols), some still embedded in the parental
clouds. Gas outflows induced by hot star winds and supernova
explosions from previous stellar generations have displaced dust and
gas (dark-gray circles) to the edges of the region. The ionized gas,
located at the boundaries of the site, `sees' the surrounding dust as
a `foreground distribution'. The galaxy's diffuse ISM (light-gray
circles) is also surrounding the starburst. Both the galactic and the
starburst-associated dust are clumpy \citep{gord97,charlot00}. Thus,
the stellar continuum light will often emerge from regions that are
not necessarily spatially coincident (in projection) with those of the
dust and ionized gas \citep{cal97}. If star formation has been
on-going for longer than the lifetime of massive stars ($>$10$^7$~yr),
the surroundings of the current starburst site become populated by the
less massive, long-lived stars from previous starburst generations
(light starred symbols), that are diffusing into the less opaque
general ISM or have dispersed their parental clouds
\citep{cal94,charlot00}. This model should be regarded as a simple
approximation of the more realistic distribution of emitters and
absorbers in starburst galaxies. \label{fig8}}

\figcaption{The UV to IR luminosity ratio as a function of the total
UV plus IR luminosity for star-forming galaxies at z$\sim$1 (panel~a)
and z$\sim$3 (panel~b). From Figure~11 of \citet{ade00}, with data
kindly provided by K. Adelberger (2001, private communication) and
modified to our selected cosmology, $\Omega_M$=1, $\Omega_{\Lambda}$=0
and H$_o$=65~km~s$^{-1}$~Mpc$^{-1}$. The (blue) circles are the
high-redshift data, the shaded region shows the locus occupied by the
local star-forming and starburst galaxies \citep{wan96,heck98}, the
(black) stars are the ULIRGs from \citet{gold01} and
\citet{tren99}. The infrared luminosity of the high redshift galaxies
is derived using UV data and the $\beta$--versus--L$_{IR}$/L$_{UV}$
correlation \citep[equation~17 and][]{ade00}. For fixed UV/IR ratio,
the faint luminosity end of the high redshift galaxy distributions is
determined by selection effects, while the bright luminosity end is
free from such effects. Thus, if UV/IR--vs.--$\beta$ relationship
holds for the distant star-forming galaxies, more actively
star-forming galaxies are on average more opaque, similar to the trend
observed in local galaxies \citep[][see also
Figure~\ref{fig5}]{wan96,heck98}. It also highlights that at fixed
UV/IR ratio (fixed opacity), higher redshift galaxies tend to be
brighter at the high luminosity end than local galaxies. Since the sum
of the UV and IR luminosities measures the SFR, high redshift galaxies
tend to be more actively star forming than local galaxies for fixed
dust opacity \citep{ade00}. \label{fig9}}

\figcaption{The attenuated-to-intrinsic L$_{H\alpha}$/L$_{0.28}$ ratio in
z$\sim$1--2 galaxies,
R$_{H\alpha,0.28}$=(L$_{H\alpha}$/L$_{0.28}$)$_{a}$/(L$_{H\alpha}$/L$_{0.28}$)$_{o}$,
as a function of the obscuration at H$\alpha$ (panel~a) and of the dust
optical depth $\tau_V$ (panel~b). The range of observational values is marked
by the two horizontal lines. The models of dust/emitter distribution are
coded as in Figure~\ref{fig3}, but only the case of the SMC extinction
curve and of the starburst obscuration curve are shown.\label{fig10}}

\figcaption{The UV-to-IR SED of a 1~Gyr constant star formation population,
attenuated by the starburst obscuration curve (equation~8), is shown for
increasing amounts of dust: E(B$-$V)$_{gas}$=0.05 (blue, continuous), 0.20
(blue dashed), 0.40 (black), 0.55 (red dashed), and 0.75 (red continuous). All
SEDs are arbitrarily normalized to the flux density at 0.17~$\mu$m. The
infrared SED is schematically represented by a single-temperature dust
component with T=50~K and $\epsilon$=2 (panel~a) and T=40~K and
$\epsilon$=1.5 (panel~b), to highlight differences in the long--wavelength
regime. \label{fig11}}





\clearpage

\begin{deluxetable}{llllll}
\tabletypesize{\scriptsize}
\tablecaption{Properties of Radiative Transfer Models\label{tbl-1}}
\tablewidth{0pt}
\tablehead{
\colhead{Dust\tablenotemark{a}} & \colhead{Medium\tablenotemark{b}} &
\colhead{Stellar\tablenotemark{c}}  &
\colhead{Reference} & \colhead{Extinction\tablenotemark{d}}   &
\colhead{Comments}  \\
\colhead{Geometry} & \colhead{Structure} &
\colhead{Distrib.}  & \colhead{   } & \colhead{Curves}   &
\colhead{      }
}
\startdata
spherical& homog. & central/  & & &\\
         &        & spherical &\citet{wit92}&MW & multiple density profiles,\\
         &        &           &             &   & multiple scale-lengths\\
         &        &           &\citet{row93}&   & starburst/IR modelling,\\
         &        &           &             &   & central source only\\
         &        &           &\citet{wis96}&   & multiple scale-lengths\\
         &        &           &\citet{varo99}&   & also external source\\
         &        &           &\citet{ferr99a}&MW,SMC& multiple dust/star density\\ 
         &        &           &             &   & profiles\\
         &        &           &\citet{efs00}&   & starburst/IR modelling, \\
         &        &           &             &   & multiple spheres, central\\
         &        &           &             &   & source only\\
         &        &           &\citet{wit00}&MW,SMC&updated albedo/phase \\ 
         &        &           &             &   & function values\\
  & & & & & \\
spherical& clumpy & central/  & & &\\
         &        & spherical &\citet{wit96} &MW &central source only\\
         &        &           &\citet{varo99}&   & also external source\\
         &        &           &\citet{wit00}&MW,SMC &updated albedo/phase \\
         &        &           &             &   & function values\\
  & & & & & \\
plane-paral.&homog.&plane-paral.&\citet{bru88}&MW,LMC &multiple scale-heights\\
         &         &            &\citet{dis89}&       &no scattering\\
         &         &            &\citet{dib95}&MW     &multiple scale-heights\\
         &         &            &\citet{xu95} &        &multiple scale-heights\\
  & & & & & \\
expon.   &homog.  &expon.     &\citet{kyl87}&MW     &edge-on only, isothermal\\
         &        &           &             &       &vertical distribution\\
         &        &           &\citet{dis89}&       &no scattering\\
         &        &           &\citet{byu94}&MW     &stellar bulge, multiple\\ 
         &        &           &             &      & scale lengths/heights\\
         &        &           &\citet{cor96}&MW     &face-on only\\
         &        &           &\citet{kuc98}&MW     &stellar bulge\\
         &        &           &\citet{ferr99a}&MW,SMC& stellar bulge, multiple\\
         &        &           &             &      & scale lengths/heights\\
  & & & & & \\
expon.   &clumpy  &expon.     &\citet{kuc98}&MW     &stellar bulge\\
         &        &           &\citet{silv98}&MW     &stellar bulge, IR\\
         &        &           &              &       &modelling\\
         &        &           &\citet{bia00}&MW     &clumping of stars as well\\
  & & & & & \\
arbitrary&homog./ & arbitrary &  & & \\
         &clumpy  &           &\citet{tre99}     &   &Multi-phase clumpy medium\\
         &        &           &              &       &and emitters\\         &        &           &\citet{gor01,mis01}&All&Multi-phase clumpy medium\\
         &        &           &              &       &and emitters, polarization,\\ 
         &        &           &              &       &IR modelling\\
\enddata


\tablenotetext{a}{The dust geometrical distribution: spherical,
plane-parallel slab, double-exponential, or arbitrary.}
\tablenotetext{b}{The structure of the dust, i.e., homogeneous or
clumpy (clumpy media can be two-phase or multi-phase).}
\tablenotetext{c}{The distribution of the stellar population: central
to the dust distribution, diffuse through the spherical dust
distribution (with variable scale length), plane-parallel,
double-exponential, or arbitrary. In most cases, a single SED is
assumed and multiple stellar populations are not included.}
\tablenotetext{d}{The adopted extinction curve(s): MW=Milky~Way;
LMC=Large Magellanic Cloud; SMC=Small Magellanic Cloud;
All=MW,LMC,SMC.}  

\end{deluxetable}

\clearpage

\begin{table}
\begin{center}
\caption{Reddening between Hydrogen Emission lines\label{tbl-2}}
\begin{tabular}{lrrr}
\tableline\tableline
Line & $\lambda_1$,$\lambda_2$  & L$_{\lambda_1}$/L$_{\lambda_2}$\tablenotemark{a} & k($\lambda_2$)$-$k($\lambda_1$)\\
Ratio & ($\mu$m,$\mu$m)         &                                   &   \\
\tableline
H$\alpha$, H$\beta$ & 0.6563,0.4861 & 2.87 & 1.163\\
H$\alpha$, H$\gamma$ & 0.6563,0.4340   & 6.16  & 1.628\\
Br$\gamma$, H$\beta$ & 2.166,0.4861 & 0.0302 & 3.352 \\
Br$\gamma$, Pa$\beta$ & 2.166,1.282 & 0.175 & 0.457\\
Pa$\alpha$, H$\alpha$ & 1.876,0.6563 & 0.123 & 2.104\\
\tableline
\end{tabular}


\tablenotetext{a}{Variations in the ratio due to variations in the gas
temperature between 5000~K and 20000~K amount to $\sim$5\%--10\%
\citep{oste89}.}

\tablecomments{The table columns list: (1) The pair of hydrogen emission
lines; (2) the vacuum wavelengths of the two lines; (3) their intrinsic
luminosity ratio for temperature T=10,000~K and case~B recombination
\citep{oste89}; (4) the differential extinction between the two wavelengths,
for a foreground non-scattering dust screen and a MW extinction curve for 
diffuse ISM \citep{ccm89}. Note that k(V)=3.1 and k(H$\alpha$)=2.468.}
\end{center}
\end{table}

\clearpage

\begin{deluxetable}{lccccccccc}
\tabletypesize{\scriptsize}
\tablecaption{Average Dust Attenuation in Local Galaxies\label{tbl-3}}
\tablewidth{0pt}
\tablehead{
\colhead{Galaxy Type} & \colhead{L$_{dust}$/L$_{bol}$\tablenotemark{a}} &
\colhead{A$_{0.15, f}$\tablenotemark{b}}  &
\colhead{A$_{0.15}$\tablenotemark{c}} &\colhead{A$_{B, f}$\tablenotemark{b}}&
\colhead{A$_B$\tablenotemark{c}} & \colhead{A$_{I, f}$\tablenotemark{b}}   &
\colhead{A$_I$\tablenotemark{c}} & \colhead{A$_{K, f}$\tablenotemark{b}}   &
\colhead{A$_K$\tablenotemark{c}} 
}
\startdata
E/S0&0.05--0.15 & 0.10--0.20 &\nodata& 0.05--0.10 &\nodata &0.02--0.04
&\nodata &$<$0.02 &\nodata\\
Sa--Sab&\nodata &0.40--0.65 &0.90--1.25&0.20--0.40&0.50--0.75 &0.10--0.15
&0.30--0.40&$\lesssim$0.05 & $\lesssim$0.15\\
Sb--Scd&0.45--0.65 &0.60--0.80 &1.20--1.45 &0.30--0.50&0.65--0.95
&0.15--0.20&0.40--0.45 & 0.05--0.10 & 0.15--0.20\\
Irr &0.25--0.40 &0.30--0.45 & 0.60--0.75 & 0.10--0.15 & 0.30--0.40 &$\sim$0.05
&$\sim$0.15 &$\lesssim$0.02 &$\lesssim$0.06\\
 \enddata


\tablenotetext{a}{Fraction of the bolometric radiation absorbed and/or
scattered by dust.}  
\tablenotetext{b}{Face-on attenuation, in magnitudes, at the specified
wavelength/band. The B-band and I-band values are derived from various
authors (see text). The other wavelengths are from model~4 for 
E/S0s, and from model~5 of section~2.2 for disks and irregulars.}
\tablenotetext{c}{Inclination-averaged attenuation for the disk and
irregular galaxies, from model~5 of section~2.2. For observational
data at 0.15~$\mu$m, see \citet{bel01}.} 

\end{deluxetable}


\clearpage

\begin{table}
\scriptsize
\begin{center}
\caption{Local UV-selected Starbursts\label{tbl-4}}
\begin{tabular}{lcrrrrrrrr}
\tableline\tableline
Name & D  & Morphology &L$_{bol}$ & SFR & L$_{dust}$/L$_{bol}$ 
& A$_{0.15}$ &A$_B$    &A$_I$     & A$_{H\alpha}$\\
     & (Mpc) &     & (L$_{\odot}$) & (M$_{\odot}$~yr$^{-1}$)&     
& (mag)     & (mag)    & (mag)    &  (mag)       \\
\tableline
\multicolumn{10}{c}{The Dust-Rich Starbursts\tablenotemark{a}}  \\
\tableline
NGC1614 &74. &SBc, merger. &4.8E+11  &54.9 &0.91 &5.67 &2.78 &1.20 &3.08\\
NGC4194 &40. &IBm, merger  &1.1E+11  &9.9  &0.86 &3.20 &1.57 &0.68 &1.74\\
NGC6090 &135. &Sd, merger  &4.1E+11  &52.4 &0.88 &3.41 &1.67 &0.72 &1.85\\
\tableline
\multicolumn{10}{c}{The Dust-Poor Starbursts\tablenotemark{a}}  \\
\tableline
NGC1140 &23. &IBm  &9.1E+09 &1.1  &0.59   &0.83 &0.40 &0.17 &0.45\\
NGC1510 &12. &SA0  &8.1E+08 &0.06 &0.42  &0.52 &0.25 &0.11 &0.28\\
Tol1924$-$416 &44. &Irr &1.9E+10 &2.8 &0.42 &0.52 &0.25 &0.11 &0.28\\
\tableline
\multicolumn{10}{c}{The Average UV-Selected Starburst\tablenotemark{a}}  \\
\tableline
        &    &     &    &   &  &1.60  &0.78 &0.34 &0.86\\
\tableline
\normalsize
\end{tabular}


\tablenotetext{a}{Examples of three dust-rich (top) and three dust-poor
(middle) starbursts in the UV-selected sample of \citet{cal94}. The
obscuration properties of these two extremes can be compared with those of an
average UV-selected starburst (last line of Table). Dustier systems are
associated with more luminous host galaxies \citep{heck98}.}

\tablecomments{The table columns give: (1) the galaxy name; (2) its distance
in Mpc, for H$_o$=65~km~s$^{-1}$~Mpc$^{-1}$; (3) the galaxy morphology; (4)
the bolometric luminosity, in solar units; (5) the star formation rate,
calculated from the obcuration-corrected Br$\gamma$ or H$\alpha$ line emission
\citep{cal97}, using the conversion SFR=8.2$\times$10$^{-40}$~L$_{Br\gamma}$
=7.9$\times$10$^{-42}$~L$_{H\alpha}$ \citep{ken98}; (6) the fraction of
bolometric luminosity re-processed by dust in the infrared; (7)--(10) the
obscuration of the 0.15~$\mu$m, B and I stellar continuum and of the H$\alpha$
emission line \citep{cal94,cal97,cal00}. For UV-selected starbursts,
A$_{H\beta}$=1.47~A$_{H\alpha}$.}
\end{center}
\end{table}

\clearpage

\begin{table}
\begin{center}
\caption{Reddening and Obscuration of z$\sim$3 Galaxies\label{tbl-5}}
\begin{tabular}{rlrrrr}
\tableline\tableline
& Parameter &SMMJ14011$+$0252& MS 1512$-$cB58 & West MMD11 & L$^*$ LBG\\
\tableline
1 &z &2.565 & 2.729 & 2.982 & $\sim$3\\
2 &AB$_{\cal{R}}$ &21.25\tablenotemark{a} & 20.41\tablenotemark{a}   & 24.05  & $\sim$24.5\\
3 &G$-\cal{R}$ &0.69\tablenotemark{b} & 0.72  & 0.80  & 0.6\\
4 &$\beta_{26}$ &$-$0.74\tablenotemark{c}& $-$1.34\tablenotemark{d} & $-$0.90\tablenotemark{e} & $-$1.5\tablenotemark{f}\\
5 & A$_{0.16, star}$ & 3.1 & 1.8 & 2.7 & 1.5 \\
6 &E(B$-$V)$_{star}$(UV) &0.31 & 0.18 & 0.27 & 0.15\\
7 &E(B$-$V)$_{star}$(H$\beta$) &\nodata &0.1$\pm$0.1\tablenotemark{g}  &0$\pm$0.2 & \nodata\\
8 &F$_{pred}$(IR) & 3.7E$-$13& 2.3E$-$13 & 1.9E$-$14 & 3.2E$-$15\\
9 &f$_{obs}$(450) &42$\pm$7 & \nodata & 22$\pm$23 & $<$10 \\
10 &f$_{pred}$(450)\tablenotemark{h} & 25.9, 71.6 & 19.5, 51.9 & 2.1, 5.3 &
0.36, 0.89\\
11 &f$_{obs}$(850) &15$\pm$2& $<$4 & 5.5$\pm$1.4 & $<$1\\
12 &f$_{pred}$(850)\tablenotemark{h} & 4.0, 19.0 & 3.2, 14.6 & 0.4, 1.6 &
0.06, 0.27\\
\tableline
\end{tabular}


\tablenotetext{a}{The AB magnitudes of SMMJ14011 and MS~1512$-$cB58 are not corrected for the lens magnification factor. AB$_{\cal{R}}$ of
MS~1512$-$cB58 is at the observer's wavelength 0.654~$\mu$m \citep{ellin96},
and has intrinsic value AB$_{\cal{R}}\simeq$24.5.}
\tablenotetext{b}{The G$-\cal{R}$ color of SMMJ14011$+$0252 has not been 
corrected for the opacity of the intervening Lyman$\alpha$ Forest 
\citep{ade00}.}
\tablenotetext{c}{As derived by \citet{ade00}.}
\tablenotetext{d}{Derived from $\beta_{18}$=$-$1.10 using
equation~B2. $\beta_{18}$ was directly measured from the UV spectrum kindly 
provided by M. Pettini \& C. Steidel (2000, private communication).}  
\tablenotetext{e}{Derived by convolving a 1~Gyr constant star
formation synthetic SED reddened by Ly$\alpha$ Forest and dust
absorption with the G and $\cal{R}$ bandpasses (moved to the galaxy's
restframe) until the observed color is reproduced, and then fitting a
power law through the UV windows of \citet{cal94}.}
\tablenotetext{f}{From Figure~12 of \citet{ade00}.}
\tablenotetext{g}{The measured H$\alpha$/H$\beta$ emission line ratio
of MS~1512$-$cB58 \citep{teplit00} yields 
E(B$-$V)$_{star}$=0.03$^{+0.10}_{-0.03}$.}
\tablenotetext{h}{Derived for dust temperature T=50~K and emissivity
$\epsilon$=2 (left hand-side value), and for T=40~K and $\epsilon$=1.5 (right 
hand-side value).}

\tablecomments{Predicted and observed values for three highly reddened
z$\sim$3 galaxies, the two lensed SMMJ14011$+$0252 and MS~1512$-$cB58, and
West~MMD11, compared with the average z$\sim$3 L$^*$ LBG. SMMJ14011$+$0252 is
SCUBA-discovered star forming galaxy \citep{ivi00,ade00}. The listed
parameters are: (1) the redshift; (2) the AB magnitude at the
observer's frame 0.693~$\mu$m (see note for MS~1512$-$cB58); (3) the
G$-\cal{R}$ color (corresponding to a 0.12$-$0.17 color for a z=3 galaxy),
corrected for the opacity of the intervening Lyman$\alpha$ Forest
\citep{pet01}; (4) the UV slope $\beta_{26}$ (Appendix~B); (5)--(6) the
stellar obscuration at 0.16~$\mu$m and the color excess derived from the UV
slope $\beta_{26}$ (section~4.3), assuming an intrinsic UV slope $\beta_{26,
0}=-$2.2, which is equivalent to a $\sim$1~Gyr continuous star formation
population; (7) the stellar color excess derived from the
L$_{0.16}$/L$_{H\beta}$ luminosity ratio \citep[for a discussion of the
uncertainties due to model assumptions see][]{pet01}; (8) the predicted
infrared flux, in erg~s$^{-1}$~cm$^{-2}$, of the dust emission, using the
IR/UV--vs.--$\beta$ correlation (equation~17); (9) and (11) the observed flux
densities at 450~$\mu$m and 850~$\mu$m, in mJy, from \citet{ivi00} for
SMMJ14011$+$0252, \citet{saw01} and \citet{van00} for MS~1512$-$cB58, and
\citet{cha00} for West~MMD11; (10) and (12) predicted flux densities at
450~$\mu$m and 850~$\mu$m, in mJy, for a single-temperature modified blackbody
(equation~20), and two choices of the dust temperature and emissivity; for
reference, the predicted 1.2~mm flux densities are 1.0~mJy (T=50~K and
$\epsilon$=2) and 5.7~mJy (T=40~K and $\epsilon$=1.5) for MS~1512$-$cB58, to
be compared with the observed value 1.06$\pm$0.35~mJy \citep{baker01}.}
\end{center}
\end{table}

\clearpage 

\begin{table}
\normalsize
\begin{center}
\caption{Ultraviolet Spectral Slopes\label{tbl-6}}
\begin{tabular}{lrr}
\tableline\tableline
Age (Myr) & $\beta^{inst}_{26}$ & $\beta^{const}_{26}$ \\
\tableline
2 &$-$2.70 &$-$2.68\\
5 &$-$2.50 &$-$2.61\\
10 &$-$2.45 &$-$2.57\\
15 &$-$2.30 &$-$2.55\\
30 &$-$2.09 &$-$2.50\\
50 &$-$1.93 &$-$2.45\\
100 &$-$1.61&$-$2.38\\
300 &$-$0.20&$-$2.28\\
500 &\nodata&$-$2.26\\
900 &\nodata&$-$2.23\\
\tableline
\normalsize
\end{tabular}


\tablecomments{The UV spectral slope $\beta_{26}$, as defined in
Appendix~B, is given for both instantanous burst (second column) and
constant star formation (third column) populations at selected ages,
for the dust-free case. The model populations
are from \citet{lei99}, for solar metallicity.}
\end{center}
\end{table}


		
\begin{figure}
\figurenum{1}
\plotone{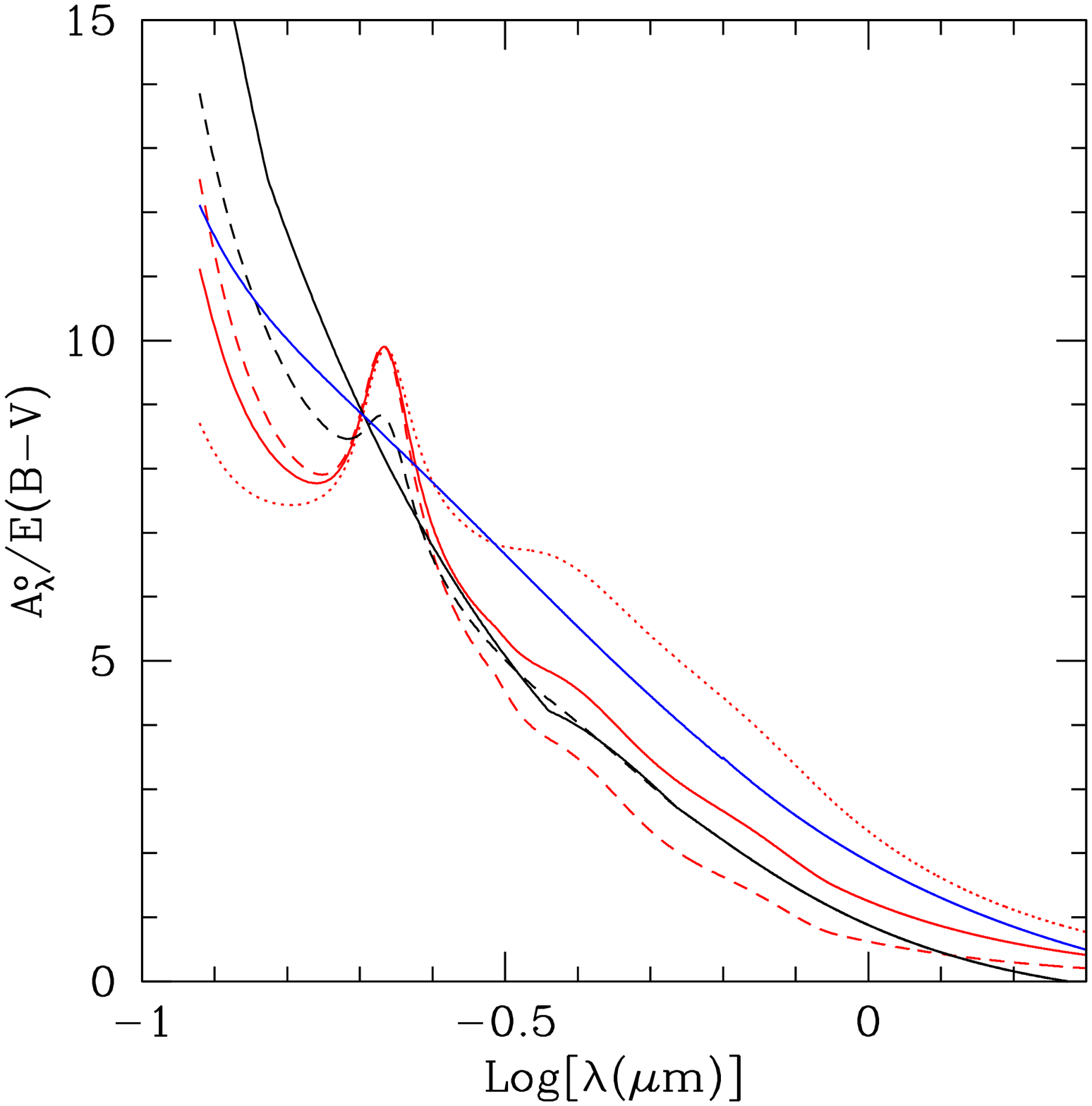}
\figcaption[figure1.eps]{}
\end{figure}

\begin{figure}
\figurenum{2}
\plotone{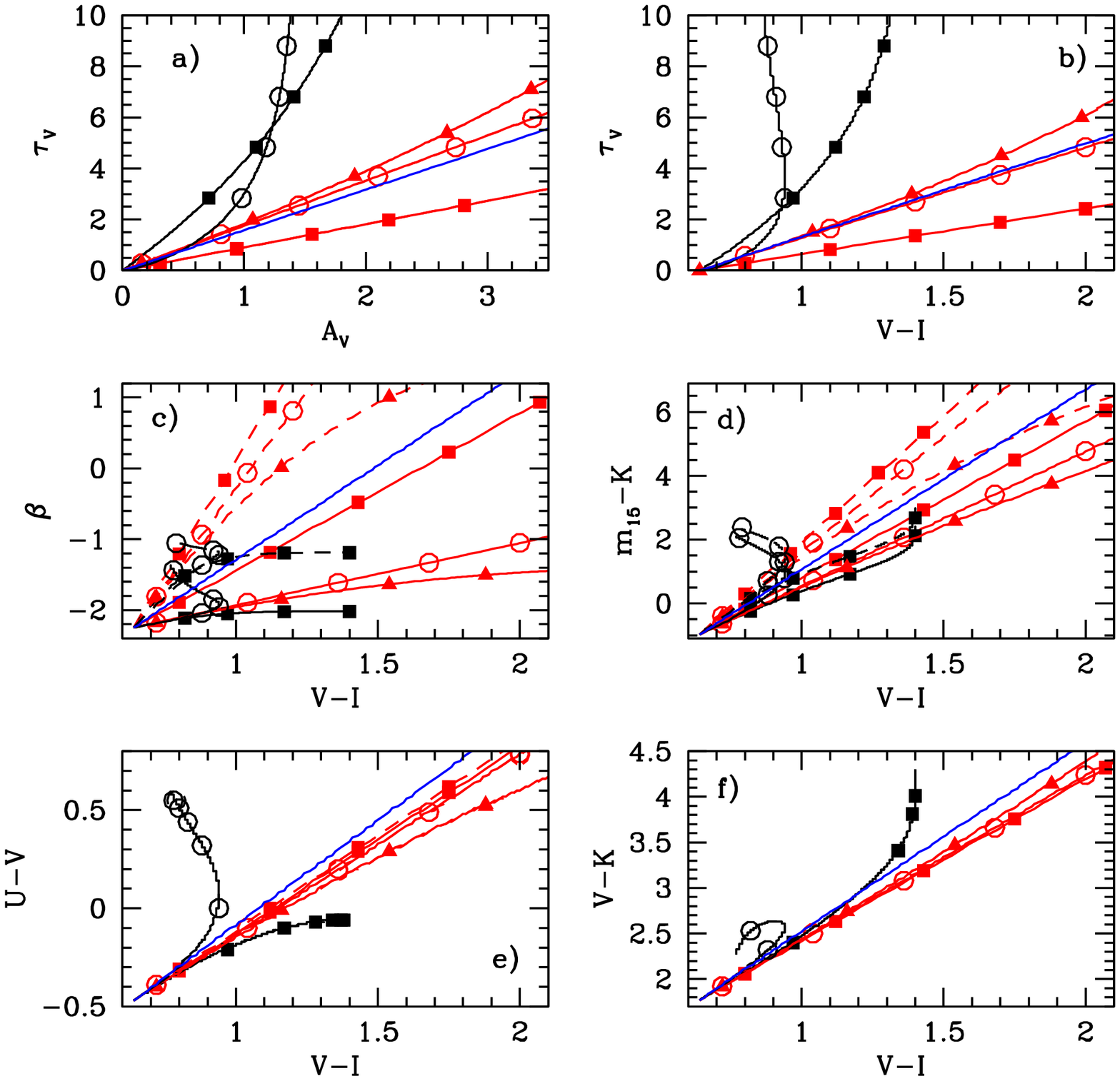}
\figcaption[figure2.eps]{}
\end{figure}

\begin{figure}
\figurenum{3}
\plotone{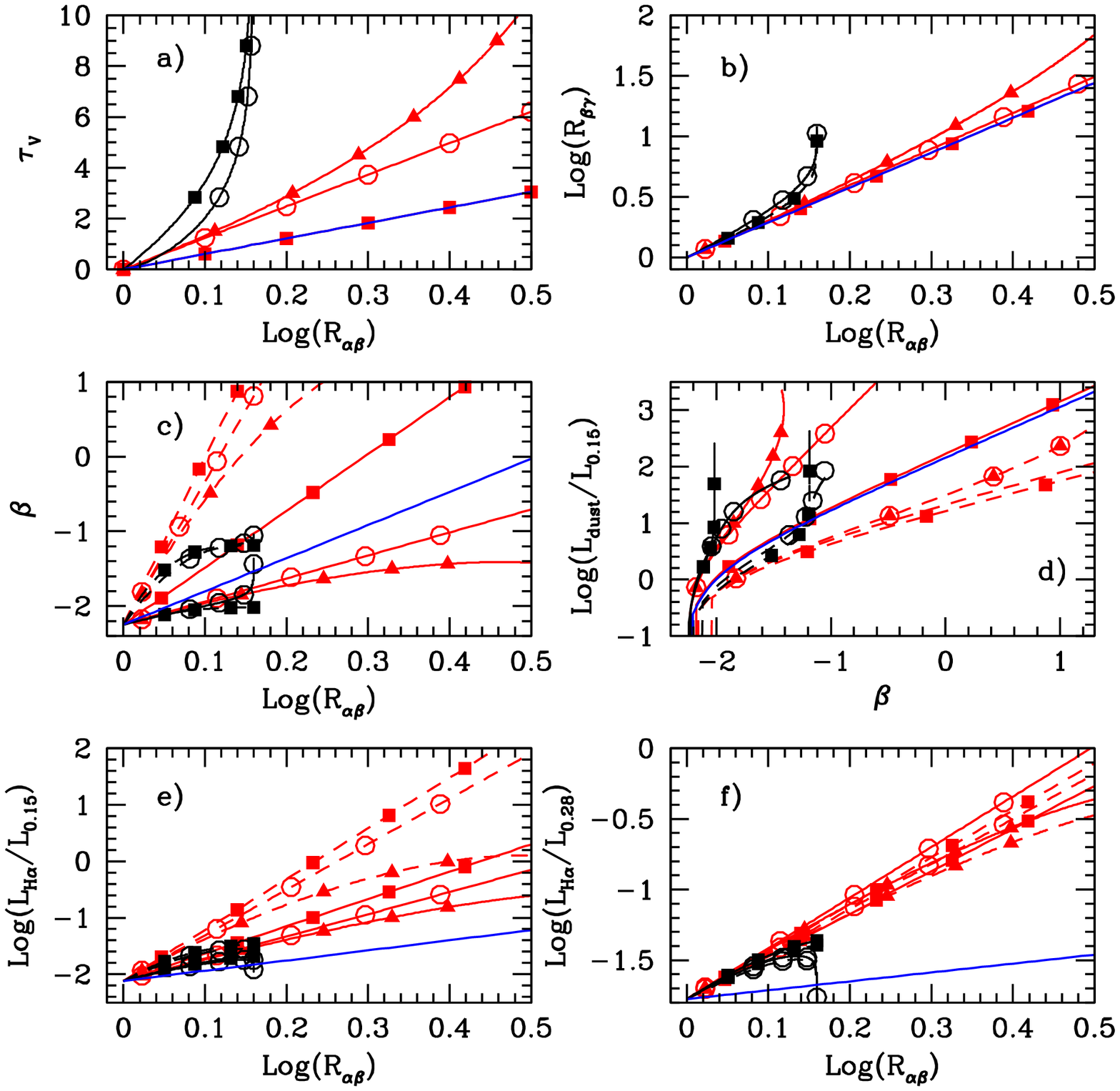}
\figcaption[figure3.eps]{}
\end{figure}

\begin{figure}
\figurenum{4}
\plotone{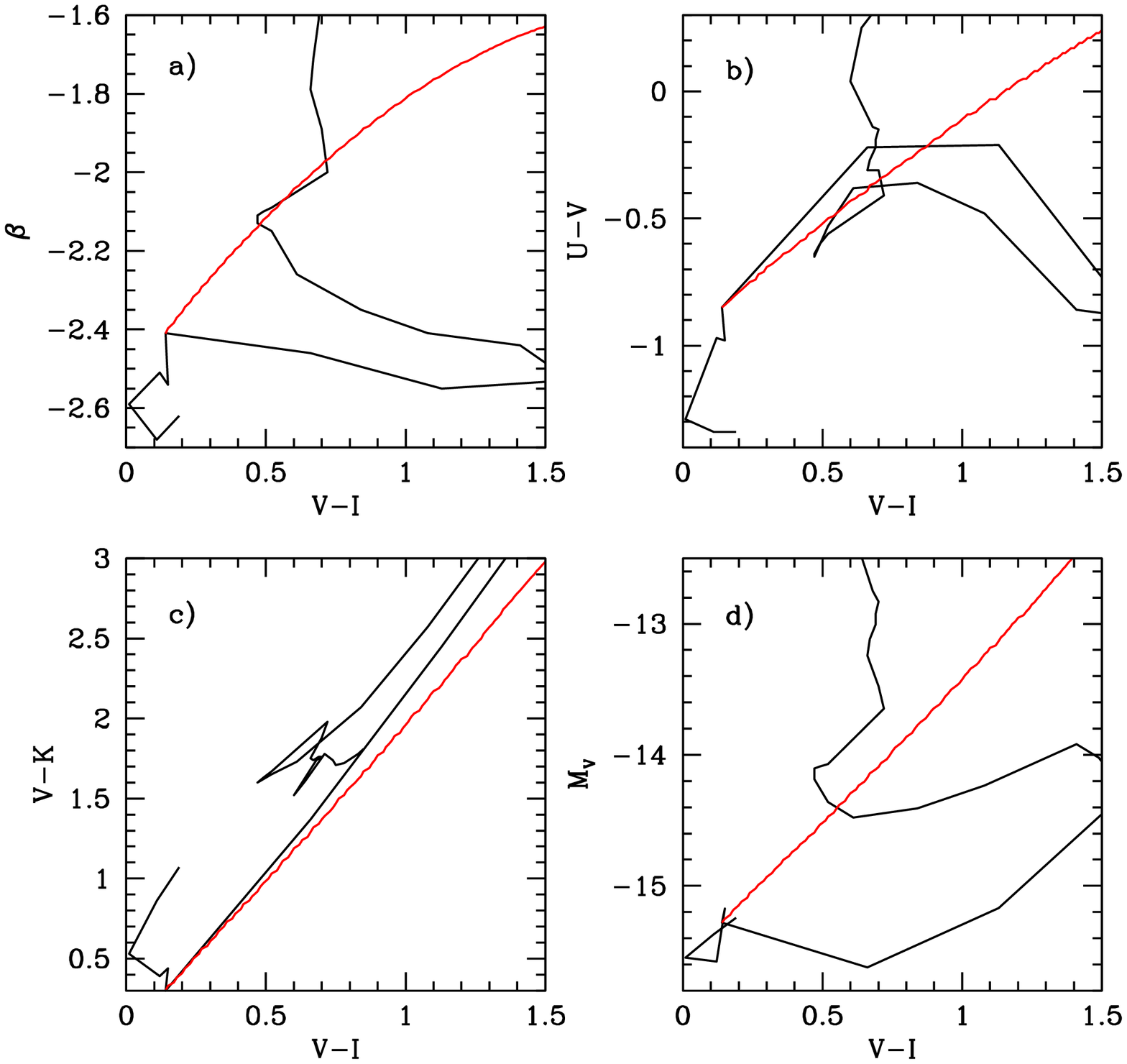}
\figcaption[figure4.eps]{}
\end{figure}

\begin{figure}
\figurenum{5}
\plotone{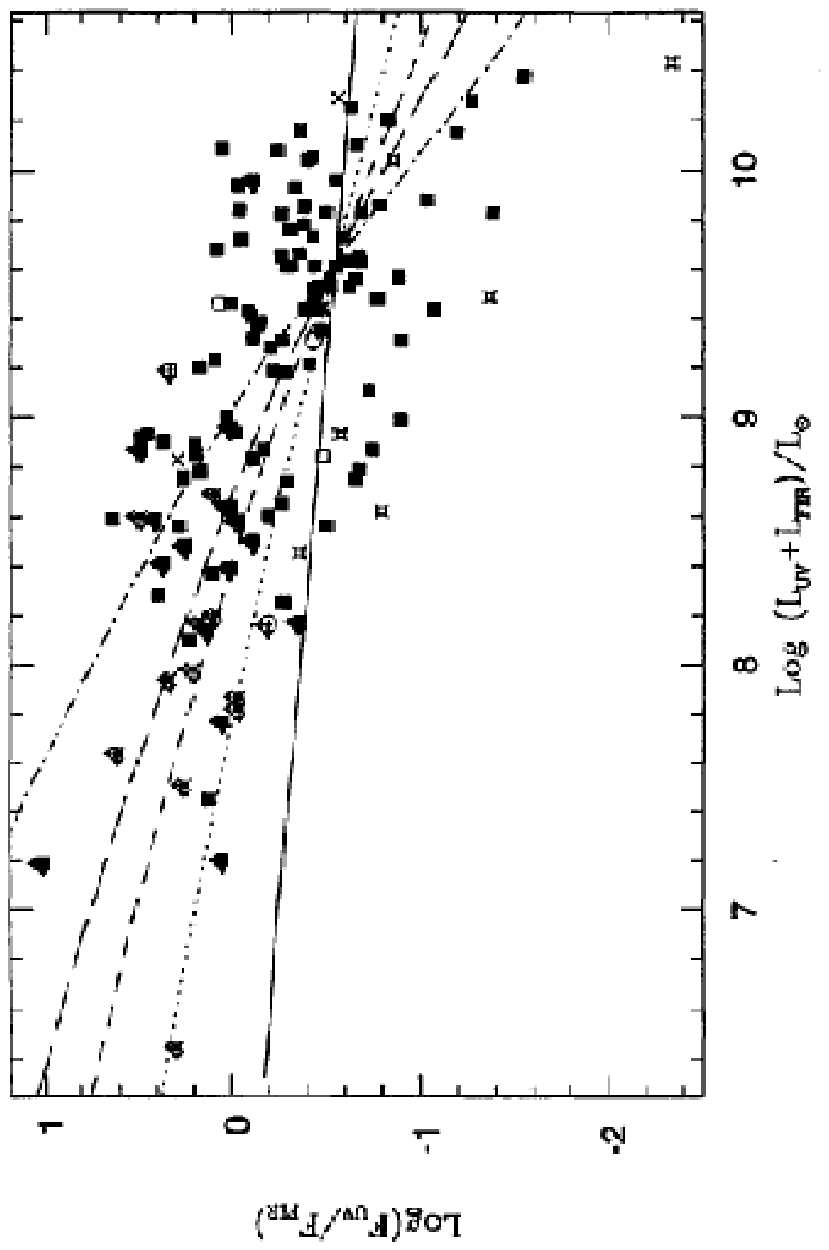}
\figcaption[figure5.eps]{}
\end{figure}

\begin{figure}
\figurenum{6}
\plotone{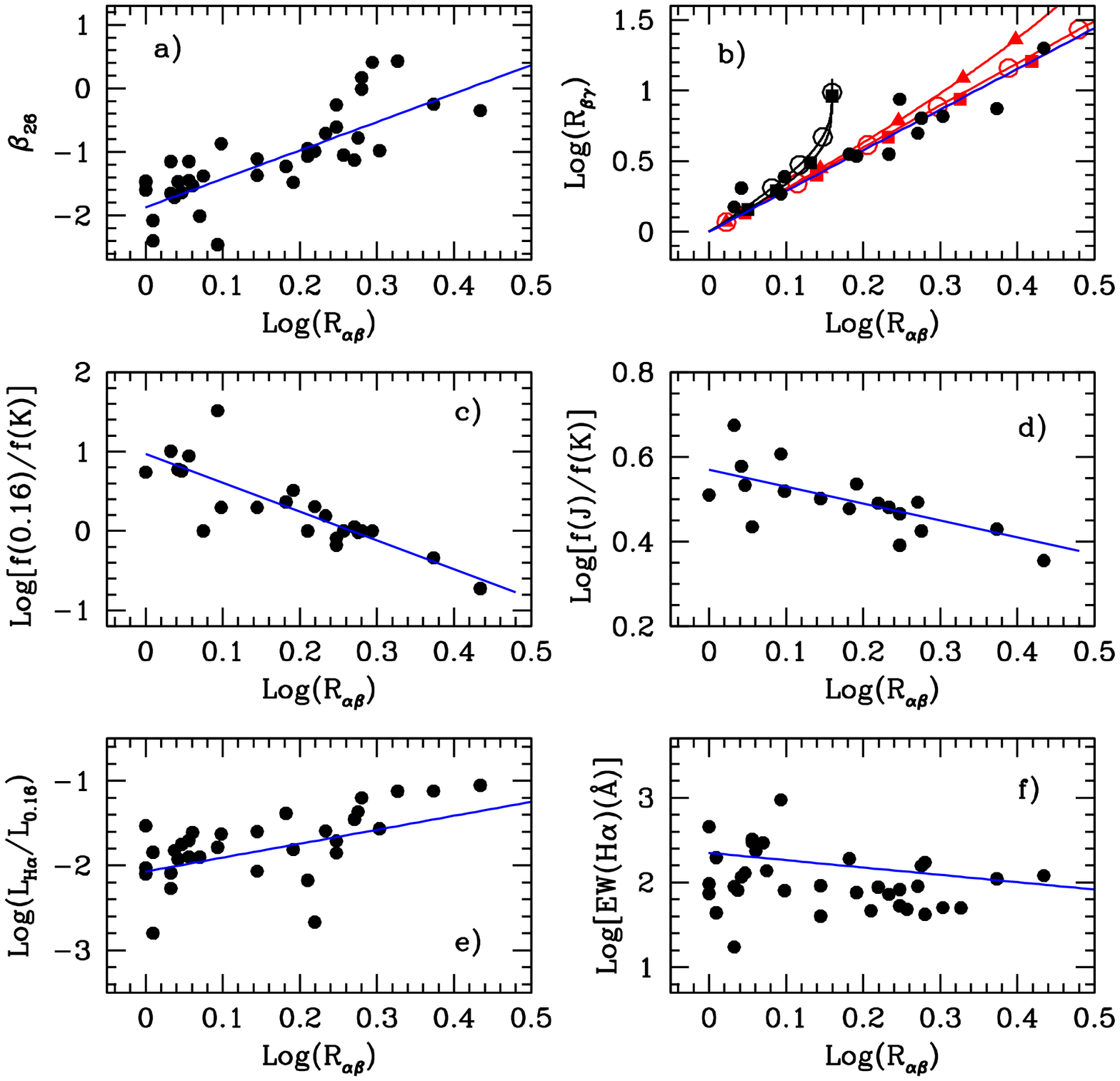}
\figcaption[figure6.eps]{}
\end{figure}

\begin{figure}
\figurenum{7}
\plotone{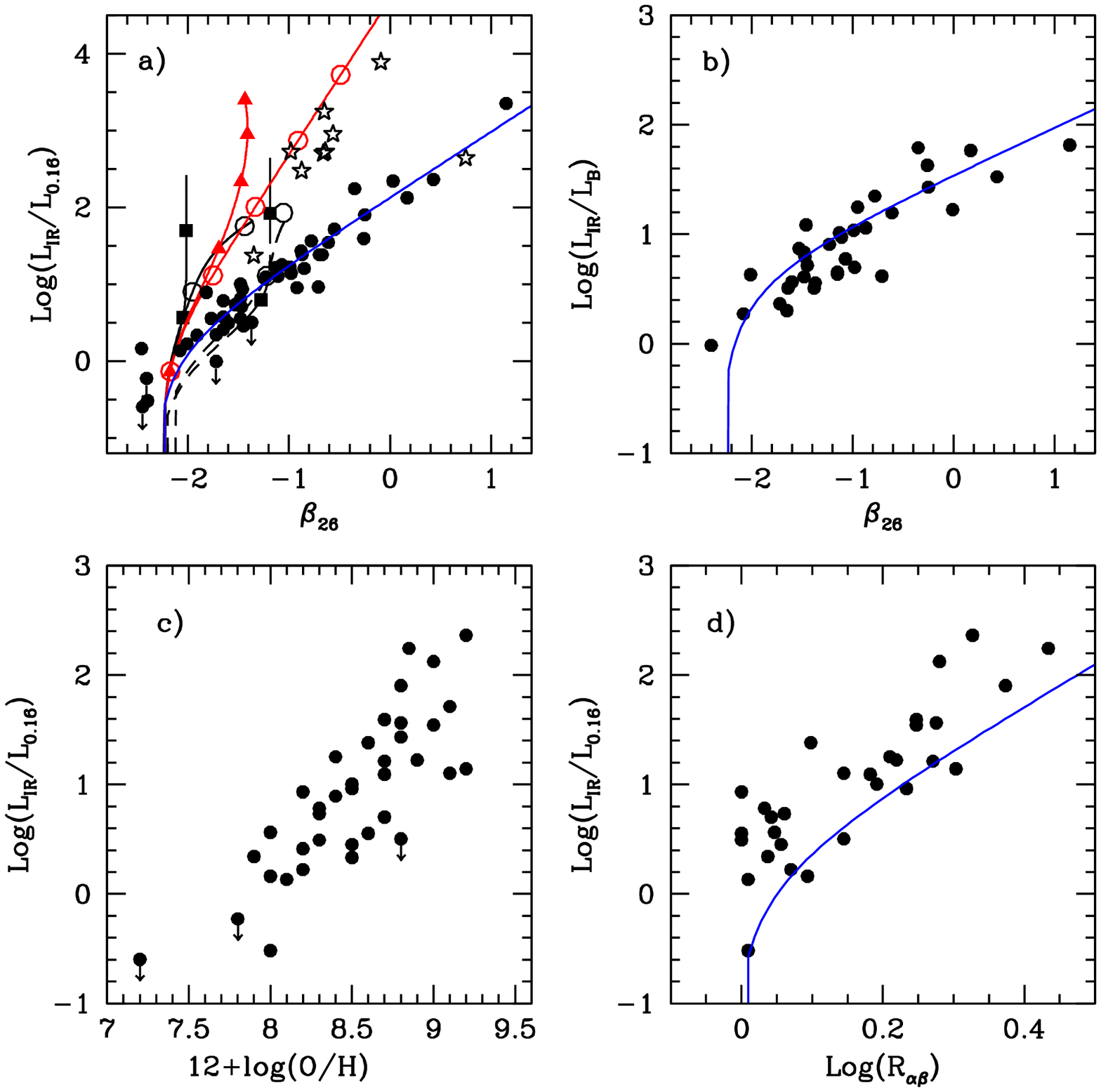}
\figcaption[figure7.eps]{}
\end{figure}

\begin{figure}
\figurenum{8}
\plotone{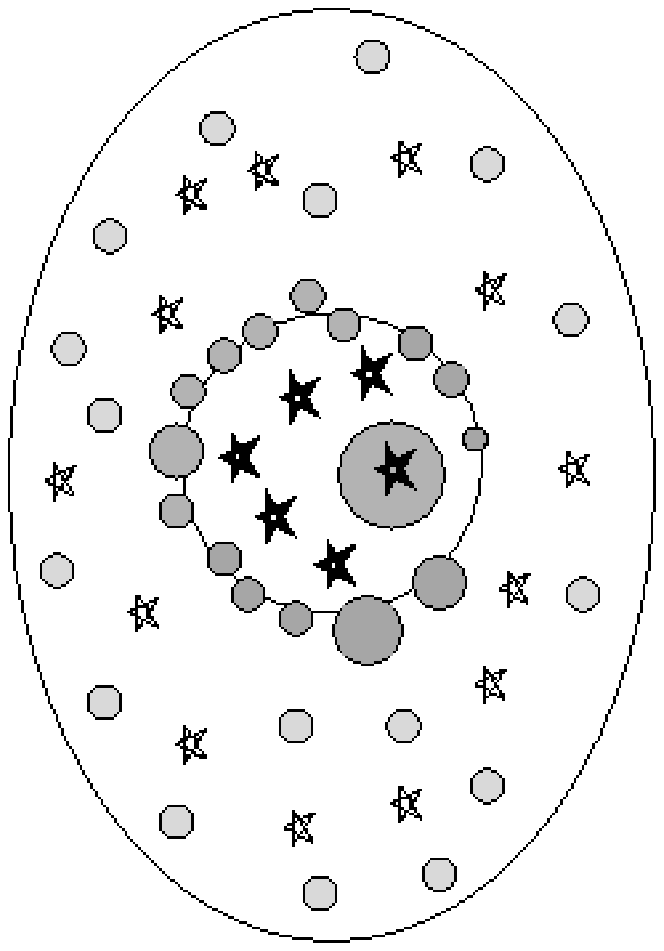}
\figcaption[figure8.eps]{}
\end{figure}

\begin{figure}
\figurenum{9a}
\plotone{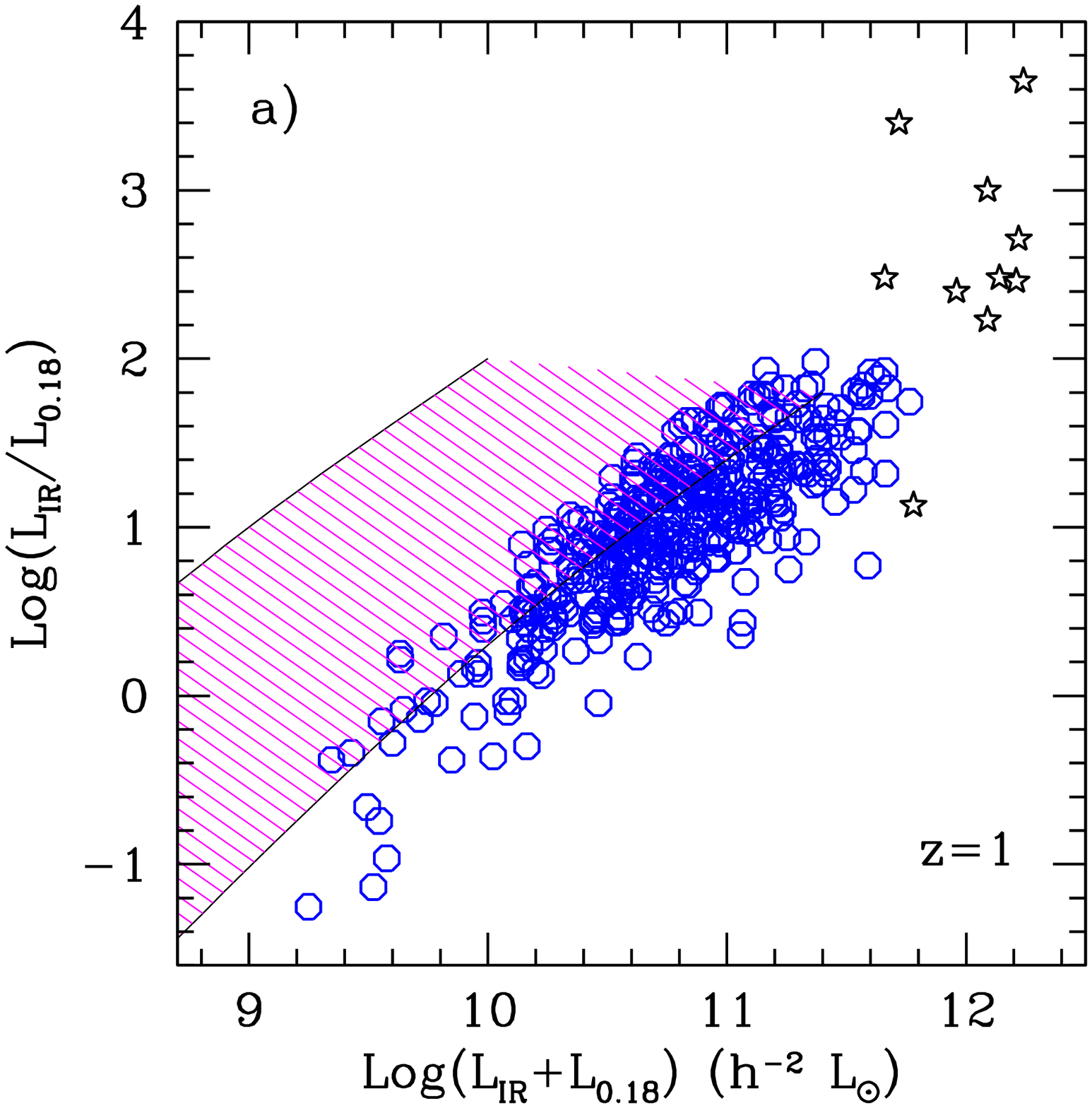}
\figcaption[figure9a.eps]{}
\end{figure}

\begin{figure}
\figurenum{9b}
\plotone{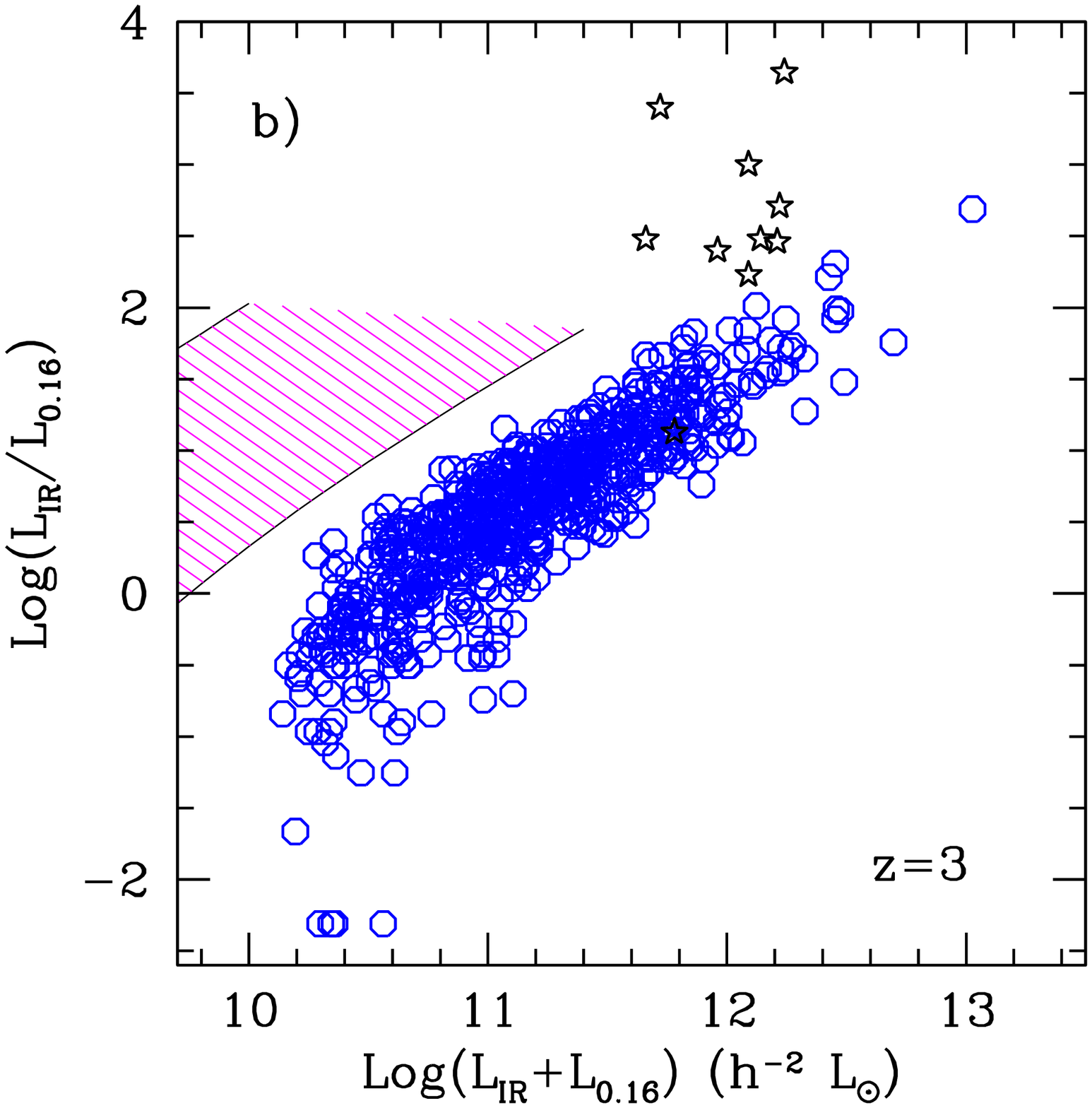}
\figcaption[figure9b.eps]{}
\end{figure}

\begin{figure}
\figurenum{10}
\plotone{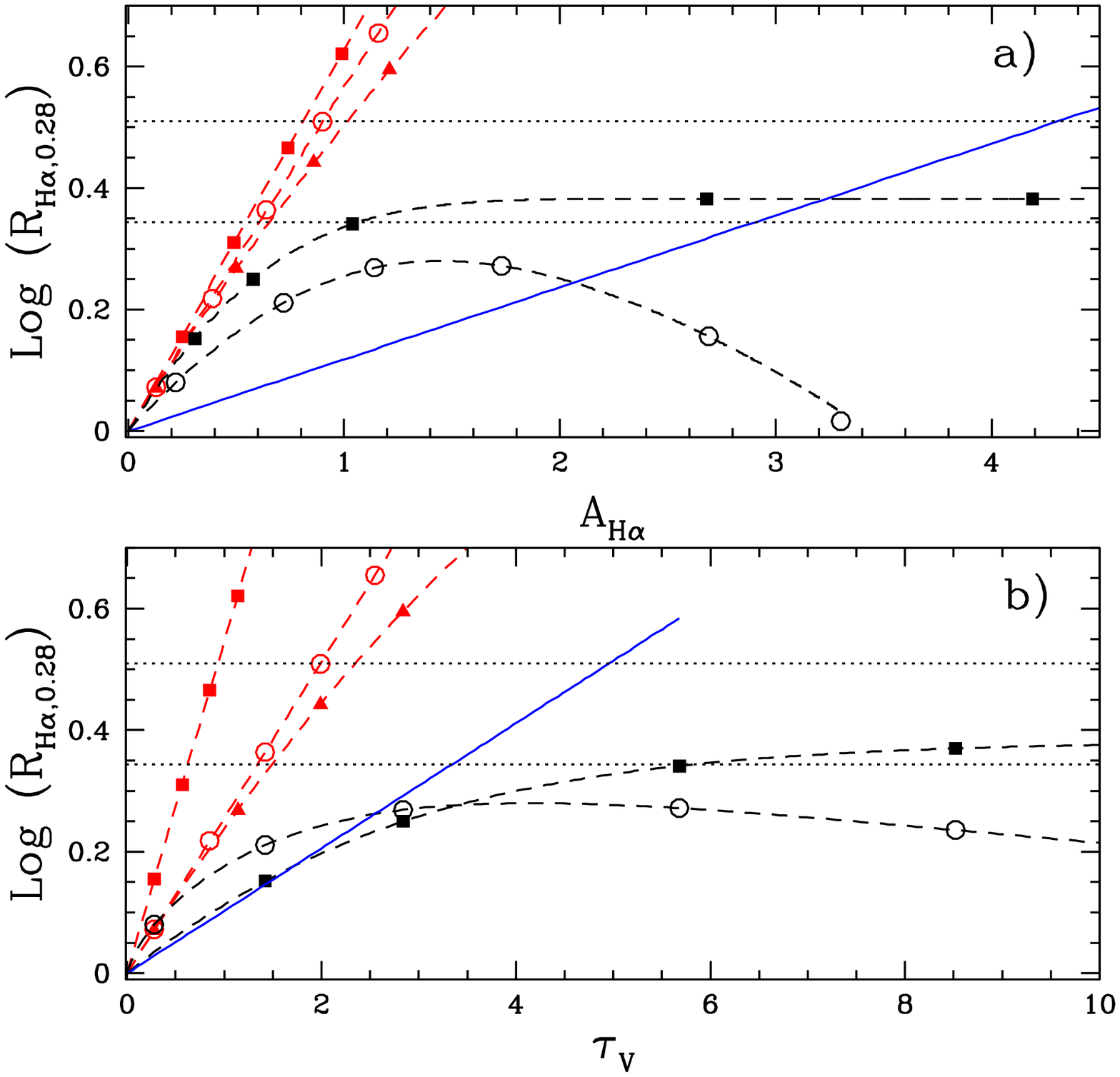}
\figcaption[figure10.eps]{}
\end{figure}

\begin{figure}
\figurenum{11}
\plotone{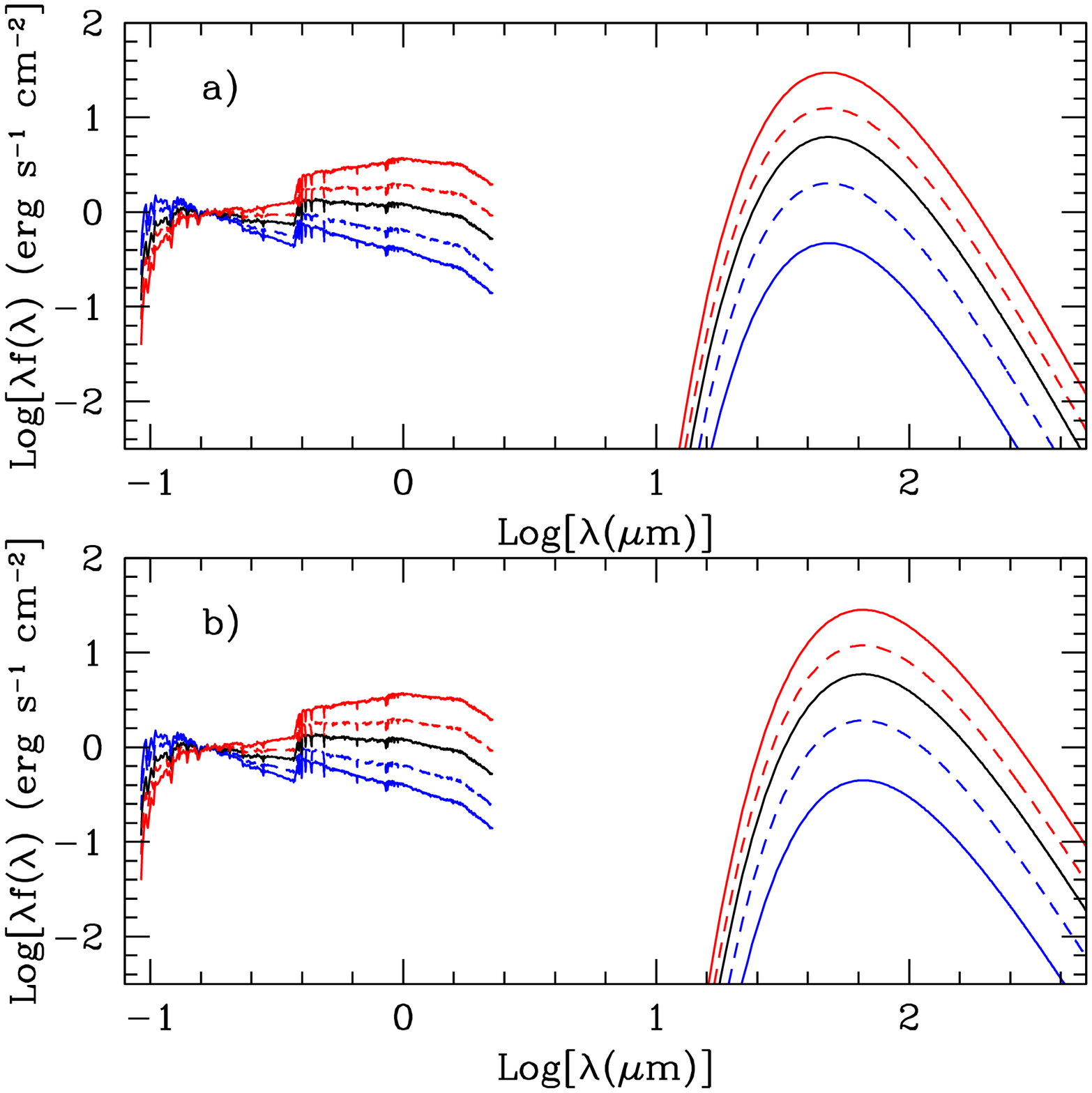}
\figcaption[figure11.eps]{}
\end{figure}


\end{document}